\documentclass[a4paper,fleqn,usenatbib]{mnras}

\usepackage{newtxtext,newtxmath}

\usepackage[T1]{fontenc}
\usepackage{ae,aecompl}
\usepackage{widetext}
\usepackage{epstopdf}
\usepackage{graphicx}	
\usepackage{amsmath}	
\usepackage{amssymb}	
\usepackage{url,bm}
\usepackage{caption}
\usepackage{subfigure}
\usepackage{CJKutf8}
\usepackage{natbib}
\usepackage{enumitem,units,multirow}
\usepackage{xcolor}
\usepackage{etoolbox}

\newcommand{\cov}{{\rm Cov}}

\newcommand{\ie}{{i.e.}}

\newcommand{\eg}{{e.g.}}

\newcommand{\ii}{{\rm i}}
\newcommand{\ee}{{\rm e}}

\title[FFT Covariance]{2D-FFTLog: Efficient computation of real space covariance matrices for galaxy clustering and weak lensing}

\author[X. Fang et al.]{
Xiao Fang (\begin{CJK*}{UTF8}{gbsn}方啸\end{CJK*}),$^{1}$\thanks{E-mail: xfang@email.arizona.edu}
Tim Eifler,$^{1}$
and Elisabeth Krause$^{1,2}$
\\
$^{1}$Department of Astronomy and Steward Observatory, University of Arizona, 933 N Cherry Ave, Tucson, AZ 85719, USA\\
$^{2}$Department of Physics, University of Arizona, 1118 E. Fourth Street, Tucson, AZ 85721, USA
}

\date{Accepted XXX. Received YYY; in original form ZZZ}

\pubyear{2020}

\begin{document}
\label{firstpage}
\pagerange{\pageref{firstpage}--\pageref{lastpage}}
   \maketitle

\begin{abstract}
Accurate covariance matrices for two-point functions are critical for inferring cosmological parameters in likelihood analyses of large-scale structure surveys. Among various approaches to obtaining the covariance, analytic computation is much faster and less noisy than estimation from data or simulations. However, the transform of covariances from Fourier space to real space involves integrals with two Bessel integrals, which are numerically slow and easily affected by numerical uncertainties. Inaccurate covariances may lead to significant errors in the inference of the cosmological parameters. In this paper, we introduce a 2D-FFTLog algorithm for efficient, accurate and numerically stable computation of non-Gaussian real space covariances for both 3D and projected statistics. The 2D-FFTLog algorithm is easily extended to perform real space bin-averaging. We apply the algorithm to the covariances for galaxy clustering and weak lensing for a Dark Energy Survey Year 3-like and a Rubin Observatory's Legacy Survey of Space and Time Year 1-like survey, and demonstrate that for both surveys, our algorithm can produce numerically stable angular bin-averaged covariances with the flat sky approximation, which are sufficiently accurate for inferring cosmological parameters. The code \textsc{CosmoCov} for computing the real space covariances with or without the flat sky approximation is released along with this paper.
\end{abstract}

\begin{keywords}
cosmology: theory -- large-scale structure of Universe -- dark energy -- cosmological parameters
\end{keywords}




\section{Introduction}
\label{sec:intro}
Ongoing and upcoming photometric and spectroscopic surveys, such as Kilo-Degree Survey (KiDS)\footnote{http://kids.strw.leidenuniv.nl}, Hyper Suprime-Cam (HSC)\footnote{https://www.naoj.org/Projects/HSC}, Dark Energy Survey (DES)\footnote{https://www.darkenergysurvey.org}, Rubin Observatory's Legacy Survey of Space and Time (LSST)\footnote{https://www.lsst.org}, Nancy Grace Roman Space Telescope\footnote{https://wfirst.gsfc.nasa.gov}, Spectro-Photometer for the History of the Universe, Epoch of Reionization, and Ices Explorer (SPHEREx)\footnote{http://spherex.caltech.edu}, Extended Baryon Oscillation Spectroscopic Survey (eBOSS)\footnote{https://www.sdss.org/surveys/eboss}, Dark Energy Spectroscopic Instrument (DESI)\footnote{https://www.desi.lbl.gov}, Subaru Prime Focus Spectrograph (PFS)\footnote{https://pfs.ipmu.jp}, Euclid\footnote{https://www.euclid-ec.org}, aim for precise measurements of cosmological parameters. While likelihood-free approaches are emerging as a tool for cosmological inference \citep[\eg,][]{2015JCAP...08..043A,2017A&A...599A..79P}, most of cosmological analyses still rely on a likelihood function $\mathcal{L}(\mathbfit{D}\vert\mathbfit{M})$, and the simplest and most common choice is a multivariate Gaussian distribution,
\begin{equation}
    \mathcal{L}(\mathbfit{D}\vert\mathbfit{M}(\mathbfit{p})) = \frac{1}{\sqrt{(2\pi)^{N_D}\vert\mathbfss{C}\vert}}\exp\left\lbrace-\frac{1}{2}[\mathbfit{M}(\mathbfit{p})-\mathbfit{D}]^{\rm T}\mathbfss{C}^{-1}[\mathbfit{M}(\mathbfit{p})-\mathbfit{D}]\right\rbrace~,
\label{eq:likelihood}
\end{equation}
where $\mathbfit{D}$ is the data vector, $\mathbfit{M}(\mathbfit{p})$ is the model prediction with parameters $\mathbfit{p}$, $\mathbfss{C}$ is the covariance matrix, and $N_D$ is the dimension of the data vector. In principle, whether one can apply the Gaussian likelihood assumption depends strongly on the observable in question, and its limitations for 2-point statistics in the context of cosmic shear are discussed in \eg, \citep{2009A&A...504..689H,2010PhRvL.105y1301S,2013A&A...551A..88C,2018MNRAS.477.4879S}. However, its impact on the cosmological parameter inferences may be insignificant especially for future surveys \citep[\eg,][]{2019arXiv190503779L}. We will only consider data vectors consisting of two-point correlation functions in real space, since many weak lensing analyses are performed in real space, such as KiDS \citep{2017MNRAS.465.1454H}, DES \citep{2018PhRvD..98d3528T,2018PhRvD..98d3526A}, and HSC \citep{2019arXiv190606041H}.

Broadly, there are three approaches to obtaining covariance matrices: estimation from simulations, estimation from data, and analytical computation. Estimation from simulations requires a large number of large, high resolution numerical simulations, due to the intrinsic noise properties of the maximum likelihood estimator of the covariance \cite{2013MNRAS.432.1928T,2013PhRvD..88f3537D,2014MNRAS.442.2728T}. This approach is computationally too expensive as the number of data points increases \citep[but see][for hybrid estimators]{2017MNRAS.466L..83J,2018MNRAS.473.4150F}. Estimation from data has the advantage of not depending on the model used in the simulations or in the analytic calculation. However, the estimated covariance is noisy due to the limited number of independent realizations available from the data, leading to the loss of information \citep{2007A&A...464..399H,2013PhRvD..88f3537D,2014MNRAS.439.2531P,2017MNRAS.464.4658S}. Moreover, because the unbiased covariance estimator of simulations or data leads to biased estimation of the precision matrix, an additional term must be introduced to the likelihood to be marginalized over in order to correct the impact on parameter inferences \citep{2016MNRAS.456L.132S}. There have also been numerous efforts made to improve the data- or simulation-based estimation of covariances from theory, or reduce the number of simulations required \citep[\eg][]{2016MNRAS.460.1567P,2019MNRAS.483..189H,2019MNRAS.487.2701O,2019MNRAS.490.5931P}. Altogether, there is a strong motivation for robust, precise methods for analytic covariances. We focus on the efficient and accurate evaluation of real space covariance matrices, assuming that a sufficiently accurate model of the Fourier space covariance exists.

The transform of a covariance matrix of 3D statistics from Fourier space to real space involves double integrals with two spherical Bessel functions in the integrand (\ie, ``double Bessel integrals''). It is numerically challenging due to the highly oscillatory nature of the spherical Bessel functions, and some recent works have proposed estimating the galaxy clustering covariance directly in real space, avoiding the Bessel function integrals \citep[\eg][]{2016MNRAS.462.2681O}. For projected angular statistics, the transform involves double summations over multipole moments weighted by combinations of Legendre and associated Legendre polynomials. Assuming narrow angles or flat sky, the transform is well approximated by double integrals with two Bessel functions in the integrand, and suffers from similar numerical challenges. We will call the former ``curved sky covariance'' and the latter ``flat sky covariance''.

In this paper, we present a new FFTLog based algorithm to efficiently compute the double Bessel integrals for the real space covariance matrices. In \S\ref{sec:algorithm}, we introduce the 2D-FFTLog algorithm for covariances and detail its implementation. The 2D-FFTLog avoids the computation of hypergeometric functions, leading to simpler expressions. It is worth noting that our method, when only considering the Gaussian covariances, is similar in nature to the method proposed in \S3 of \cite{2019JCAP...01..016L}.

Furthermore, covariance matrices are often averaged over each real space (or angular) bin since the correlation functions are measured in bins \citep[\eg,][]{2017MNRAS.471.3827S}. We present an accurate way of bin-averaging in our 2D-FFTLog algorithm. In \S\ref{sec:3by2pt}, we apply our algorithm to analytic covariance matrices for the joint-probe analysis of galaxy clustering, galaxy-galaxy lensing (GGL), and cosmic shear. In \S\ref{sec:validation}, we compare the flat sky FFT covariances to the flat sky covariances from the traditional quadrature (Quad) integration and the curved sky covariances in the context of future surveys such as DES Year 3 (DES Y3) and LSST Year 1 (LSST Y1). We also perform simulated likelihood analyses and show that the flat sky covariances from the 2D-FFTLog algorithm do not introduce biases in the cosmological parameters for these surveys. Finally, we discuss other applications and summarize in \S\ref{sec:discussion}.

We emphasize that the method presented in this paper can be directly applied to future surveys including DES, LSST, and Nancy Grace Roman Space Telescope. Its advantage in efficiency will be even more prominent when more tomographic bins are used, and when more probes (such as CMB lensing, cluster clustering and cluster lensing) are jointly analyzed.

\section{2D-FFTLog for Covariances}
\label{sec:algorithm}
The 2D-FFTLog algorithm presented here is a generalization of the original FFTLog, which has found many applications in cosmology. In \S\ref{subsec:motiv}, we first introduce the transforms of covariances from Fourier space to real space and motivate the use of the 2D-FFTLog algorithm. In \S\ref{subsec:NG}, we introduce the algorithm for the general non-Gaussian covariance integral Eq.~(\ref{eq:covNG}). We also show in \S\ref{subsec:G} that the Gaussian covariance integral Eq.~(\ref{eq:covG}) can be evaluated with the same algorithm. In \S\ref{subsec:binave} we present a simple way to perform real space bin-averaging.

\subsection{Motivation for 2D-FFTLog}\label{subsec:motiv}
For covariances of 3D statistics, the numerically challenging transform takes the form of double Bessel integrals\footnote{The form is only for observables based on two-point correlators. Furthermore, it requires the correlators to be functions of $r$ only, \ie, it does not apply to full 3D correlation functions $\xi(r, \mu)$, though it should work for the multipoles \citep[\eg][]{2016MNRAS.457.1577G}.}
\begin{equation}
C(r_1,r_2) = \int_0^\infty \frac{dk_1}{k_1}\int_0^\infty \frac{dk_2}{k_2} f(k_1,k_2) j_{\scriptscriptstyle L_1}(k_1 r_1)j_{\scriptscriptstyle L_2}(k_2 r_2)~,
\label{eq:covNG}
\end{equation}
while the Gaussian covariances take the form of
\begin{equation}
C_{\scriptscriptstyle\rm G}(r_1,r_2) = \int_0^\infty \frac{dk}{k} f(k) j_{\scriptscriptstyle L_1}(k r_1)j_{\scriptscriptstyle L_2}(kr_2)~.
\label{eq:covG}
\end{equation}
For given types of statistics (two-point correlation functions), $L_1,L_2$ are fixed. The functions $f(k)$ and $f(k_1,k_2)$ are smooth functions of $k$ or $k_1,k_2$. To evaluate all the elements of an $N_r\times N_r$ covariance matrix, the quadrature integration of Eq.~(\ref{eq:covG}) will take order of $N_r^2 N_k$ steps, where $N_k$ is the number of $k$ points sampled, while Eq.~(\ref{eq:covNG}) will take order of $N_r^2 N_k^2$ steps. Furthermore, the rapidly oscillatory Bessel functions, possibly beating with each other, require very small integration step sizes (\ie, very large $N_k$), especially when $r_1,r_2$ are large.

For flat sky covariances of projected statistics, the spherical Bessel functions $j_{\scriptscriptstyle L}$'s are replaced by Bessel functions, which can be converted back to this form using $J_n(z)=\sqrt{2z/\pi}j_{n-1/2}(z)$ (see \S\ref{sec:3by2pt} for relevant discussions). The curved sky covariances are detail in \S\ref{subsubsec:covs}.

Originally developed for atomic physics applications, the FFTLog algorithm \citep{1978JCoPh..29...35T,2000MNRAS.312..257H} has been used to efficiently perform Fourier transforms with logarithmic variables, and single Bessel integrals, \ie, the Hankel transforms, in the form of $\int_0^\infty dk f(k)j_{\scriptscriptstyle L}(kr)$ or $\int_0^\infty dk f(k)J_{n}(kr)$, with logarithmically sampled $k$. The core idea of the FFTLog algorithm involves a power-law decomposition, in the form of $f(k)=\sum_m c_m k^{z_m}$ ($z_m$ may be complex), so that each term in the decomposition can be integrated analytically. The FFTLog has found many applications in cosmology, including the one-loop order \citep{2016PhRvD..93j3528S,2016JCAP...09..015M,2017JCAP...02..030F} and two-loop order nonlinear perturbation theories \citep{2016PhRvD..94j3530S,2018JCAP...04..030S,2018arXiv181202728S}, the angular power spectra, bispectra \citep{2017JCAP...11..054A,2018PhRvD..97b3504G,2018arXiv180709540S,2019arXiv191111947F,2019arXiv191200065S} and trispectra \citep{2020arXiv200100584L}, the real space Gaussian covariances \citep{2019arXiv191200065S}, and the Fourier space non-Gaussian covariances \citep{2020arXiv200100584L}.

In particular, the angular power spectra and bispectra without the Limber approximation contain double Bessel integrals in the form of Eq.~(\ref{eq:covG}). The same power-law decomposition leads to double Bessel transforms of power laws, which have analytical solutions containing Gauss hypergeometric functions \citep[as shown in][]{2017JCAP...11..054A,2018PhRvD..97b3504G,2018arXiv180709540S}. We refer to this method as the 1D-FFTLog method. Note that the 1D-FFTLog method cannot be applied to the non-Gaussian covariances Eq.~(\ref{eq:covNG}). Hypergeometric functions can be numerically challenging to evaluate, and require specialized methods to improve the speed and stability \citep{2017JCAP...11..054A,2018PhRvD..97b3504G,2018arXiv180709540S}. \cite{2019arXiv191111947F} presents a new method for efficiently computing non-Limber angular power spectra with a generalized FFTLog algorithm without invoking hypergeometric functions, but it is not applicable to covariances. 

\cite{2019arXiv191200065S} introduces a ``rotation method'', which decomposes the spherical Bessel functions into series of products of polynomials and sine/cosine functions, and computes them with the FFTLog algorithm. This method is especially useful for computing Gaussian covariances of 3D statistics and possibly the non-Limber angular power spectra, at least at low $L$'s. However, it may not be applied to non-Gaussian covariances, as well as covariances of projected statistics involving $J_n$'s which cannot be written in terms of sine/cosine functions.

All the methods mentioned above cannot be applied to the general form of the real space non-Gaussian covariances (Eq.~\ref{eq:covNG}, with integers or half-integers $L_1,L_2$). Since non-Gaussian covariances generally take more time to evaluate than Gaussian covariances do, we are motivated to develop a new FFTLog algorithm for efficiently solving the double Bessel integrals.

\subsection{Transform for Non-Gaussian Covariances}\label{subsec:NG}
We decompose the discretized two-variable smooth function $f(k_1,k_2)$ in Eq.~(\ref{eq:covNG}) into a series of products of two power-laws, which is equivalent to the 2-dimensional Fourier series in the $\log k_1$-$\log k_2$ space. Assuming that the $k_1,k_2$ arrays are identical and sampled logarithmically (from now on, denoting $k_i$ as the $i$-th element of $k_1$ or $k_2$ array, for $i=0,1,\cdots,N-1$), we have
\begin{equation}
    f(k_{p},k_{q}) = \frac{1}{N^2}\sum_{m=-N/2}^{N/2}\sum_{n=-N/2}^{N/2} \tilde{c}_{mn} k_{0}^{-\ii\eta_m}k_{0}^{-\ii\eta_n} k_{p}^{\nu_1+\ii\eta_m}k_{q}^{\nu_2+\ii\eta_n}~,
\end{equation}
where coefficients $\tilde{c}_{mn}$ are given by a discrete Fourier transform\footnote{We use the standard convention for the discrete Fourier tranform, \ie, with minus sign in the exponent and no prefactor in front of the summation. The inverse Fouier transform has a positive sign in the exponent and a $1/N$ prefactor (or $1/N^2$ for the 2D case).}
\begin{equation}
    \tilde{c}_{mn}=\sum_{p=0}^{N-1}\sum_{q=0}^{N-1}\frac{f(k_p,k_q)}{k_{p}^{\nu_1}k_{q}^{\nu_2}}\ee^{-2\pi\ii (mp+nq)/N}~,
\label{eq:tilde_c_mn}
\end{equation}
$N$ is the size of the $k$ array. $\eta_m=2\pi m/(N\Delta_{\ln k})$, $\nu_1,\nu_2$ are the real parts of the power law indices (which we call ``bias parameters''), and $\Delta_{\ln k}$ is the linear spacing in $\ln k$, \eg, $k_{q} = k_{0}\exp(q\Delta_{\ln k})$ with $k_{0}$ being the smallest value in the $k_1$ array. For a real function $f(k_1,k_2)$, the Fourier coefficients obey $\tilde{c}_{mn}^* = \tilde{c}_{-m,-n}$.

In practice, we apply a filter on $\tilde{c}_{mn}$ to smooth along both axes at large $m,n$ by a one-dimensional window function $\mathbfit{W}$,\footnote{We use the 1D window function $\mathbfit{W}$ described in Appendix C of \cite{2016JCAP...09..015M}.} such that $c_{mn} = W_mW_n\tilde{c}_{mn}$, where $W_m,W_n$ are the $m$-th and $n$-th elements of $\mathbfit{W}$. We then use the smoothed coefficients $c_{mn}$ for the remaining computation. This filtering process effectively removes sharp edges at the boundary of $c_{mn}$ and improves the numerical performance of the subsequent summations \citep[see][for its use in 1D]{2016JCAP...09..015M,2017JCAP...02..030F}.

With the double-power-law decomposition, Eq.~(\ref{eq:covNG}) becomes
\begin{align}
    C(r_1,r_2) =&\frac{1}{N^2}\sum_{m,n=-N/2}^{N/2} c_{mn} k_{0}^{-\ii(\eta_m+\eta_n)} \nonumber\\
    &\int_0^\infty \frac{dk_1}{k_1}\int_0^\infty \frac{dk_2}{k_2}  k_1^{\nu_1+\ii\eta_m}k_2^{\nu_2+\ii\eta_n} j_{\scriptscriptstyle L_1}(k_1 r_1)j_{\scriptscriptstyle L_2}(k_2r_2)~\nonumber\\
    =&\frac{1}{N^2}\sum_{m,n=-N/2}^{N/2} c_{mn} k_{0}^{-\ii(\eta_m+\eta_n)}\nonumber\\
    &\int_0^\infty \frac{dk_1}{k_1} k_1^{\nu_1+\ii\eta_m} j_{\scriptscriptstyle L_1}(k_1r_1)\int_0^\infty \frac{dk_2}{k_2}k_2^{\nu_2+\ii\eta_n}j_{\scriptscriptstyle L_2}(k_2r_2) \nonumber\\
	=&\frac{\pi}{16 r_1^{\nu_1}r_2^{\nu_2}}\frac{1}{N^2}\sum_{m,n=-N/2}^{N/2} c_{mn} k_{0}^{-\ii(\eta_m+\eta_n)} r_1^{-\ii\eta_m}r_2^{-\ii\eta_n} \nonumber\\
	&\qquad\qquad\qquad\qquad\times g_{\scriptscriptstyle L_1}(\nu_1+\ii\eta_m)g_{\scriptscriptstyle L_2}(\nu_2+\ii\eta_n)~,
\label{eq:covNG_2dfft}
\end{align}
where in the third equality we use identities of Bessel functions\footnote{$\int_0^\infty t^\mu J_\nu(t)\,dt = 2^\mu\Gamma[(\nu+\mu+1)/2]/\Gamma[(\nu-\mu+1)/2]$ for $\Re(\mu+\nu)>-1$ and $\Re(\mu)<1/2$ \citep[Eq.~11.4.16 of][]{1972hmfw.book.....A}, and $j_n(z)=\sqrt{\pi/(2z)}J_{n+1/2}(z)$.} and define function $g_{\scriptscriptstyle L}(z)$ as
\begin{equation}
    g_{\scriptscriptstyle L}(z) = 2^z \frac{\Gamma\left(\frac{L+z}{2}\right)}{\Gamma\left(\frac{3+L-z}{2}\right)}~,~~-L<\Re(z)<2~.
\end{equation}
The valid ranges of ``bias parameters'' $\nu_1,\nu_2$ are thus $-L_1<\nu_1<2$ and $-L_2<\nu_2<2$, within which the bias parameters can be chosen arbitrarily, although different values can cause different levels of ringing effects at edges. In addition, integer values of $\nu_1,\nu_2$ should be avoided to avoid hitting singularities of the Gamma function. For convenience, we set $\nu_1=\nu_2=1.01$. In contrast to 1D-FFTLog algorithms, the 2D-FFTLog does not need Gauss hypergeometric functions.

In principle, the $r_1,r_2$ arrays are independent of the $k$ array. For a general choice of $r_1,r_2$ sampling, Eq.~(\ref{eq:covNG_2dfft}) can be computed with complexity of order $\mathcal{O}(N_k^2 N_r^2)$, where $N_k=N$. Although it looks similar to the complexity of the direct quadrature integration, $N_k$ required here is usually much smaller than that needed in the quadrature integration, as the Fourier space covariances are much smoother than the Bessel integrals.

For logarithmic sampling of $r_1,r_2$, we may take full advantage of the FFTLog algorithm and further simplify the calculation. We assume that $r_1,r_2$ are identical arrays (shorthanded as $r$ array), logarithmically sampled with linear spacing $\Delta_{\ln r}$ in $\ln r$, and set $\Delta_{\ln r}=\Delta_{\ln k}$. Therefore, the last summation in Eq.~(\ref{eq:covNG_2dfft}) can be written as
\begin{align}
        C(r_{p},r_{q}) =  \frac{\pi}{16 r_{p}^{\nu_1}r_{q}^{\nu_2}}{\rm IFFT2}[&c_{mn}^* (k_0r_{0})^{\ii(\eta_m+\eta_n)}\nonumber\\
        & \times g_{\scriptscriptstyle L_1}(\nu_1-\ii\eta_m)g_{\scriptscriptstyle L_2}(\nu_2-\ii\eta_n)]~,
\label{eq:cov_r_ifft2}
\end{align}
where $r_{p},r_{q}$ ($p,q=0,1,\cdots,N-1$) are the $p$-th and $q$-th elements in the $r$ array, respectively. IFFT2 stands for the two-dimensional Inverse Fast Fourier Transform.

The 2D-FFTLog method uses 2D FFT twice. Thus, it is an $\mathcal{O}(N^2\log N)$ algorithm. Note that by performing the inverse FFT, we obtain an $N\times N$ real space covariance matrix for the $N$ logarithmically sampled $r$ values. In practice, such a fine-grid matrix of step size $\Delta_{\ln r}$ is not needed, and one may choose to calculate the desired smaller-size covariance directly from the $c_{mn}$ matrix without using the inverse FFT, \ie, directly summing over $m,n$ in the last line of Eq.~(\ref{eq:covNG_2dfft}) for each covariance element.Whether the latter option is more optimal depends on the size of the desired matrix. In practice, bin-averaging is performed on covariance matrices, which acts as smoothing and needs less sampling points, hence computation time, for a given accuracy requirement. We discuss our bin-averaging algorithm in \S\ref{subsec:binave}.

\subsubsection{Special Case: Gaussian Covariance Integrals}
\label{subsec:G}
The Gaussian covariance $C_{\scriptscriptstyle\rm G}$ can be considered as a special case of the non-Gaussian covariance when $f(k_1,k_2)$ contains a Dirac delta function $\delta_D(k_1-k_2)$, which makes the separation of $k_1$ and $k_2$ difficult. However, in the quadrature integration, the integral is discretized, the Dirac delta is converted to the Kronecker delta, making the separation trivial.

Writing $C_{\scriptscriptstyle\rm G}$ (Eq.~\ref{eq:covG}) in a discrete sum \citep[similar to][]{2019JCAP...01..016L}
\begin{align}
    C_{\scriptscriptstyle\rm G} &=\lim_{\Delta_{\ln k}\rightarrow 0}\Delta_{\ln k} \sum_{p} f(k_{p})j_{\scriptscriptstyle L_1}(k_{p}r_1)j_{\scriptscriptstyle L_2}(k_{p}r_2)\nonumber\\
    &=\lim_{\Delta_{\ln k}\rightarrow 0}\Delta_{\ln k} \sum_{p}\sum_{q} f(k_{p})j_{\scriptscriptstyle L_1}(k_{p}r_1)j_{\scriptscriptstyle L_2}(k_{q}r_2)\delta_{pq}\nonumber\\
    &=\lim_{\Delta_{\ln k}\rightarrow 0}\Delta_{\ln k} \sum_{p}\Delta_{\ln k}\sum_{q} \frac{f(k_{p})\delta_{pq}}{\Delta_{\ln k}}j_{\scriptscriptstyle L_1}(k_{p}r_1)j_{\scriptscriptstyle L_2}(k_{q}r_2)~,
\end{align}
we find that it is a special case of the quadrature integration of the non-Gaussian case Eq.~(\ref{eq:covNG}) when we take $k_1,k_2$ as the same array and take the input function $f(k_1,k_2)$ as a diagonal matrix related to the input $N$-array $f(q)$, \ie, $f(k_1,k_2) = \Delta_{\ln k}^{-1}{\rm diag}\lbrace f(q_1),f(q_2),\cdots,f(q_{N-1})\rbrace$. Hence, it can be evaluated with the same 2D-FFTLog method in \S\ref{subsec:NG}.
Further speed-up can be achieved by realizing that the 2D discrete Fourier transform of a diagonal matrix is a circulant matrix \citep[also see][]{2019JCAP...01..016L}. Therefore, one can (1D) Fourier transform the $N$-array $q_i^{-\nu_1-\nu_2}f(q_i)/\Delta_{\ln k}$ to obtain the first row of $\tilde{c}_{mn}$ (which is only an $\mathcal{O}(N\log N)$ algorithm), and then apply $N-1$ cyclic permutations to the first row, to generate the remaining $N-1$ rows. In the context of non-Gaussian covariances, this optimization is negligible, since the total runtime will be dominated by the non-Gaussian computation which scales as $\mathcal{O}(N^2\log N)$.

Despite the above derivation requiring a discretization and an approximation that the quadrature integration step sizes are sufficiently small, the formula can be shown to be valid in general. In the continuous case, the two-variable function $f(k_1,k_2)$ can be written as $f(k_1,k_2)=k_2f(k_2)\delta_D(k_1-k_2)=f(k_2)\delta_D(\ln k_1-\ln k_2)$. Defining variables $x_1=\ln k_1/\Delta_{\ln k}, x_2=\ln k_2/\Delta_{\ln k}$, Eq.~(\ref{eq:tilde_c_mn}) can be rewritten as a continuous Fourier transform from the $(x_1,x_2)$ space to the $(y_1,y_2)$ space,
\begin{align}
    c(y_1,y_2)&= \int_{-\infty}^\infty\int_{-\infty}^\infty dx_1 dx_2 \frac{f(k_1,k_2)}{k_1^{\nu_1} k_2^{\nu_2}}{\rm e}^{-2\pi\ii (x_1y_1+x_2y_2)}\nonumber\\
    &=\int_{-\infty}^\infty\int_{-\infty}^\infty dx_1 dx_2 \frac{f(k_2)\delta_D(x_1-x_2)}{\Delta_{\ln k} k_1^{\nu_1} k_2^{\nu_2}}{\rm e}^{-2\pi\ii (x_1y_1+x_2y_2)}\nonumber\\
    &=\int_{-\infty}^\infty dx_1  \frac{f(k_1)}{\Delta_{\ln k} k_1^{\nu_1+\nu_2}}{\rm e}^{-2\pi\ii x_1(y_1+y_2)}\nonumber\\
    &=\mathcal{F}\left[\frac{f(k)/\Delta_{\ln k}}{k^{\nu_1+\nu_2}}\right](y_1+y_2)~,
\label{eq:c_y1y2}
\end{align}
where $\mathcal{F}[.]$ is the 1D Fourier transform. In the discrete form ($\tilde{c}_{mn}$), Eq.~(\ref{eq:c_y1y2}) is equivalent to the 2D discrete Fourier transform of a diagonal matrix with diagonal elements $q_i^{-\nu_1-\nu_2}f(q_i)/\Delta_{\ln k}$ as previously shown. The $(y_1+y_2)$-dependence translates to the $(m+n)$-dependence, showing that the resulting $\tilde{c}_{mn}$ matrix is circulant.

The 2D-FFTLog algorithm is sub-optimal for Gaussian covariances alone comparing to 1D-FFTLog, which scales as $\mathcal{O}(NN_r\log N)$, typically with $N>N_r$. However, when the total covariance (with both Gaussian and non-Gaussian contributions) is computed, the $c_{mn}$ from both covariance contributions may be combined to eliminate the additional computation time from the Gaussian covariances. Thus, for this case, using a consistent 2D-FFTLog method for both Gaussian and non-Gaussian covariances is optimal.

\subsection{Bin Averaging}\label{subsec:binave}
Since correlation function measurements are averaged over $r$ or $\theta$ bins of finite width, the same bin average should be applied to the covariance. The bin-averaged covariance can be written as
\begin{equation}
       {\rm Cov}(\bar{r}_i,\bar{r}_j) =\frac{\int_{\bar{r}_{i,\rm min}}^{\bar{r}_{i,\rm max}}dr_1\int_{\bar{r}_{j,\rm min}}^{\bar{r}_{j,\rm max}}\,dr_2\,r_1^{D-1} r_2^{D-1}{\rm Cov}(r_1,r_2)}{\int_{\bar{r}_{i,\rm min}}^{\bar{r}_{i,\rm max}}dr_1\,r_1^{D-1}\,\int_{\bar{r}_{j,\rm min}}^{\bar{r}_{j,\rm max}}\,dr_2\,r_2^{D-1}}~,
\label{eq:cov_binave_def}
\end{equation}
where $D$ is the dimension of the statistics (\ie, $D=2$ for projected statistics, and $D=3$ for 3D statistics), $\bar{r}_i,\bar{r}_j$ denote the $i$-th and $j$-th bins, and the subscripts $_{\rm min}$ and $_{\rm max}$ denote the bin edges. We restrict the discussion to the $D=3$ case for the rest of this section, and leave the $D=2$ case for the next section. We also assume a logarithmically uniform binning, \ie, with a constant logarithmic bin width. This assumption results in a simple expression similar to Eq.~(\ref{eq:cov_r_ifft2}) that can be evaluated with IFFT2.

Substituting Eq.~(\ref{eq:covNG_2dfft}) into Eq.~(\ref{eq:cov_binave_def}) and abbreviating the denominator of Eq.~(\ref{eq:cov_binave_def}) as $A$, we obtain
\begin{widetext}
\begin{align}
           {\rm Cov}(\bar{r}_i,\bar{r}_j)
           =&\frac{\pi}{16AN^2}\sum_{m,n=-N/2}^{N/2}c_{mn} k_{0}^{-\ii(\eta_m+\eta_n)}  g_{\scriptscriptstyle L_1}(\nu_1+\ii\eta_m)g_{\scriptscriptstyle L_2}(\nu_2+\ii\eta_n)\int_{\bar{r}_{i,\rm min}}^{\bar{r}_{i,\rm max}}r_1^{-\ii\eta_m-\nu_1+D-1}dr_1\int_{\bar{r}_{j,\rm min}}^{\bar{r}_{j,\rm max}}r_2^{-\ii\eta_n-\nu_2+D-1}dr_2\nonumber\\
           =&\frac{\pi}{16AN^2}\sum_{m,n=-N/2}^{N/2}c_{mn} k_{0}^{-\ii(\eta_m+\eta_n)}  g_{\scriptscriptstyle L_1}(\nu_1+\ii\eta_m)g_{\scriptscriptstyle L_2}(\nu_2+\ii\eta_n)\bar{r}_{i,\rm min}^{-\ii\eta_m-\nu_1+D}\bar{r}_{j,\rm min}^{-\ii\eta_n-\nu_2+D}\nonumber\\
           &\qquad\qquad\qquad\times\frac{[(\bar{r}_{i,\rm max}/\bar{r}_{i,\rm min})^{-\ii\eta_m-\nu_1+D}-1]}{D-\nu_1-\ii\eta_m}\frac{[(\bar{r}_{j,\rm max}/\bar{r}_{j,\rm min})^{-\ii\eta_n-\nu_2+D}-1]}{D-\nu_2-\ii\eta_n}\label{eq:cov_binave_general}\\
           =&\frac{\pi\bar{r}_{i,\rm min}^{\,\,D-\nu_1}\bar{r}_{j,\rm min}^{\,\,D-\nu_2}}{16AN^2}\sum_{m,n=-N/2}^{N/2}c_{mn} k_{0}^{-\ii(\eta_m+\eta_n)}\bar{r}_{i,\rm min}^{-\ii\eta_m}\bar{r}_{j,\rm min}^{-\ii\eta_n}\,g_{\scriptscriptstyle L_1}(\nu_1+\ii\eta_m)g_{\scriptscriptstyle L_2}(\nu_2+\ii\eta_n)s(D-\nu_1-\ii\eta_m,\lambda)s(D-\nu_2-\ii\eta_n,\lambda)~,
\label{eq:cov_binave}
\end{align}
\end{widetext}
\noindent where we define $\lambda$ as the logarithmic bin width, \ie, $\bar{r}_{i,\rm max}/\bar{r}_{i,\rm min}=\ee^\lambda$ for every bin, and
\begin{align}
    &s(z,\lambda)= \frac{\ee^{\lambda z}-1}{z}~.\\
    &A=\int_{\bar{r}_{i,\rm min}}^{\bar{r}_{i,\rm max}}dr_1\,r_1^{D-1}\,\int_{\bar{r}_{j,\rm min}}^{\bar{r}_{j,\rm max}}\,dr_2\,r_2^{D-1}=\bar{r}_{i,\rm min}^{\,\,D}\bar{r}_{j,\rm min}^{\,\,D}[s(D,\lambda)]^2~.
\end{align}
Thus, to accurately compute the bin-averaged covariance, we only need to multiply a few functions $s$ before applying the 2D IFFT, \ie,
\begin{align}
    &{\rm Cov}(\bar{r}_i,\bar{r}_j) = \frac{\pi}{16\bar{r}_{i,\rm min}^{\,\,\nu_1}\bar{r}_{j,\rm min}^{\,\,\nu_2}}{\rm IFFT2}\left[c_{mn}^* (k_0r_0)^{\ii(\eta_m+\eta_n)} \right.\nonumber\\
    &\times\left.g_{\scriptscriptstyle L_1}(\nu_1-\ii\eta_m)g_{\scriptscriptstyle L_2}(\nu_2-\ii\eta_n)\frac{s(D-\nu_1-\ii\eta_m,\lambda)s(D-\nu_2-\ii\eta_n,\lambda)}{[s(D,\lambda)]^2}\right]~.
\label{eq:cov_binave_final}
\end{align}
Whether performing the IFFT2 algorithm is more optimal than performing the direct summation in Eq.~(\ref{eq:cov_binave}) depends on the size of the desired matrix and the number of sampling in $k$, as discussed in \S\ref{subsec:NG}. In Eq.~(\ref{eq:cov_binave_final}), we have assumed a logarithmically uniform binning. For more general choice of binning, \eg, linear binning, we must directly compute Eq.~(\ref{eq:cov_binave_general}), which has complexity of order $\mathcal{O}(N^2 N_r^2)$.

The above equation works for the covariances of 3D correlation functions. Bin-averaged covariances of projected correlation functions (in the flat sky limit) require a slight modification due to the additional power $r^{1/2}$ from the conversion between the Bessel functions and the spherical Bessel functions (see Eqs.~\ref{eq:cov_binave_2d_def} and \ref{eq:cov_binave_2d}).

Note that starting with a finely sampled $k$ array, IFFT2 will return an equally finely sampled covariance (since we define $\Delta_{\ln r}=\Delta_{\ln k}$). Qualitatively, the numerical accuracy improves with increasing sampling points given the integration range. Since bin-averaging acts as a smoothing operation, high accuracy can be achieved with less fine sampling compared to the non-bin-averaged case. The actual number of sampling points needed depends on the function to be transformed, the order of Bessel functions and the range and spacing of the $r$ bins. We recommend to test over increasing sampling points until the results converge within the required accuracy. Analogous to the Fourier transform, FFTLog requires the function to be log-periodic to avoid ringing. Thus, we also recommend to zero-pad the function $f(k_1,k_2)$ to alleviate the ringing effects.

\section{Application to Projected Statistics: 3$\times$2pt Covariance}
\label{sec:3by2pt}

The $3\times 2$pt analysis, combining cosmic shear, galaxy-galaxy lensing, and galaxy clustering, has become a standard combination of probes in DES and LSST.

Following the notations in \cite{2017arXiv170609359K}, in the flat sky limit, we write the real space covariance of two angular two-point functions $\Xi,\Theta\in\lbrace w,\gamma_t,\xi_+,\xi_-\rbrace$ as
\begin{align}
    &{\rm Cov}(\Xi^{ij}(\theta_1),\Theta^{km}(\theta_2))
    \nonumber\\
    &=\frac{1}{4\pi^2}\int\frac{d\ell_1}{\ell_1}\int\frac{d\ell_2}{\ell_2}\ell_1^2\ell_2^2 J_{n(\Xi)}(\ell_1\theta_1)J_{n(\Theta)}(\ell_2\theta_2)\nonumber\\
    &\quad\times\left[{\rm Cov}^{\scriptscriptstyle\rm G}\left(C_\Xi^{ij}(\ell_1),C_\Theta^{km}(\ell_2)\right) + {\rm Cov}^{\scriptscriptstyle\rm NG}\left(C_\Xi^{ij}(\ell_1),C_\Theta^{km}(\ell_2)\right)\right]~,
\label{eq:3by2cov}
\end{align}
where $w$ is the galaxy angular clustering correlation, $\gamma_t$ is the galaxy-galaxy lensing two-point correlation, $\xi_{+/-}$ are the cosmic shear two-point correlations. The angular power spectra $C_{\xi_{+/-}}\equiv C^{ee}$, $C_{\gamma_t}\equiv C^{{\rm g}e}$, $C_{w}\equiv C^{{\rm g}{\rm g}}$, denoting the correlations between galaxy densities $\rm g$ and galaxy shapes $e$, are detailed in \S\ref{subsubsec:2ptfuncs}. The order of the Bessel function is given by $n=0$ for $\xi_+$ and $w$, $n=2$ for $\gamma_t$, and $n=4$ for $\xi_-$. $\theta_1,\theta_2$ are angular separations, $\ell_1,\ell_2$ are the angular wave numbers, and $i,j,k,m$ specify the tomography bins. The non-Gaussian covariance Cov$^{\scriptscriptstyle\rm NG}$ consists of a connected four-point correlation contribution \citep[\eg,][]{2002PhR...372....1C,2009MNRAS.395.2065T} and a super sample contribution \cite[\eg,][]{2017MNRAS.470.2100K,2019A&A...624A..61L}. The Gaussian covariance Cov$^{\scriptscriptstyle\rm G}$ \citep{2004PhRvD..70d3009H} contains a Kronecker delta function $\delta_{\ell_1\ell_2}$ which reduces the dimension of the integral. The implementation of the Fourier space covariance matrices is detailed in the Appendix of \cite{2017MNRAS.470.2100K}.

\paragraph*{Connection to 2D-FFTLog}
The integral Eq.~(\ref{eq:3by2cov}) can be written in the form of Eq.~(\ref{eq:covNG}) by realizing that $J_n(z)=\sqrt{2z/\pi}j_{n-1/2}(z)$, \ie, the integral has the form
\begin{align}
   &{\rm Cov}(\Xi^{ij}(\theta_1),\Theta^{km}(\theta_2)) \nonumber\\
   &= \sqrt{\theta_1\theta_2}\int\frac{d\ell_1}{\ell_1}\int\frac{d\ell_2}{\ell_2} j_{n(\Xi)-\frac{1}{2}}(\ell_1\theta_1)j_{n(\Theta)-\frac{1}{2}}(\ell_2\theta_2)f(\ell_1,\ell_2)~,
\label{eq:3x2pt_bessel}
\end{align}
where $f(\ell_1,\ell_2)$ is given by
\begin{align}
    f(\ell_1,\ell_2)= \frac{1}{2\pi^3}\ell_1^{5/2}\ell_2^{5/2}&\left[{\rm Cov}^{\scriptscriptstyle\rm G}\left(C_\Xi^{ij}(\ell_1),C_\Theta^{km}(\ell_2)\right)\right.\nonumber\\
    &\ \left.+ {\rm Cov}^{\scriptscriptstyle\rm NG}\left(C_\Xi^{ij}(\ell_1),C_\Theta^{km}(\ell_2)\right)\right]
\end{align}
This integral is now ready for our 2D-FFTLog method.

For better numerical accuracy and stability, we calculate the Gaussian and non-Gaussian parts separately. In addition, the Gaussian part integrand usually contains terms like $[C_{AC}(\ell_1)+N_A][C_{BD}(\ell_2)+N_B]$, where $C$ are angular power spectra of two random fields specified by the subscripts and $N$ are noise terms. The pure noise term $N_A N_B$, if nonzero, is taken out of the integral because the corresponding term is analytically solvable and by doing so, the double Bessel integral is numerically more stable.

\paragraph*{Bin Averaging in Flat Sky Limit}
The covariance is averaged in each angular bin, \ie,
\begin{equation}
       {\rm Cov}(\bar{\theta}_i,\bar{\theta}_j) =\frac{\int_{\bar{\theta}_{i,\rm min}}^{\bar{\theta}_{i,\rm max}}d\theta_1\int_{\bar{\theta}_{j,\rm min}}^{\bar{\theta}_{j,\rm max}}\,d\theta_2\,\theta_1\theta_2{\rm Cov}(\theta_1,\theta_2)}{\int_{\bar{\theta}_{i,\rm min}}^{\bar{\theta}_{i,\rm max}}d\theta_1\,\theta_1\,\int_{\bar{\theta}_{j,\rm min}}^{\bar{\theta}_{j,\rm max}}\,d\theta_2\,\theta_2}~,
\label{eq:cov_binave_2d_def}
\end{equation}
where $\bar{\theta}_i,\bar{\theta}_j$ are $i$-th and $j$-th angular bins, and the subscripts $_{\rm min}$ and $_{\rm max}$ denote the lower and upper bin edges. However, we cannot directly apply Eq.~(\ref{eq:cov_binave_final}) with $D=2$ due to the additional $\sqrt{\theta_1\theta_2}$ factor in Eq.~(\ref{eq:3x2pt_bessel}). Following the derivation in \S\ref{subsec:binave}, we obtain the modified bin-averaging result
\begin{widetext} 
\begin{align}
    {\rm Cov}(\bar{\theta}_i,\bar{\theta}_j)
    =&\frac{\pi}{16B N^2}\sum_{m,n=-N/2}^{N/2}c_{mn} \ell_{0}^{-\ii(\eta_m+\eta_n)}  g_{n_1-\frac{1}{2}}(\nu_1+\ii\eta_m)g_{n_2-\frac{1}{2}}(\nu_2+\ii\eta_n)\bar{\theta}_{i,\rm min}^{-\ii\eta_m-\nu_1+5/2}\bar{\theta}_{j,\rm min}^{-\ii\eta_n-\nu_2+5/2}\nonumber\\
    &\qquad\qquad\qquad\times\frac{[(\bar{\theta}_{i,\rm max}/\bar{\theta}_{i,\rm min})^{-\ii\eta_m-\nu_1+5/2}-1]}{\frac{5}{2}-\nu_1-\ii\eta_m}\frac{[(\bar{\theta}_{j,\rm max}/\bar{\theta}_{j,\rm min})^{-\ii\eta_n-\nu_2+5/2}-1]}{\frac{5}{2}-\nu_2-\ii\eta_n}\label{eq:cov_binave_2d_general}\\
    =& \frac{\pi}{16\bar{\theta}_{i,\rm min}^{\,\,\nu_1-1/2}\bar{\theta}_{j,\rm min}^{\,\,\nu_2-1/2}}{\rm IFFT2}\left[c_{mn}^* (\ell_0\theta_0)^{\ii(\eta_m+\eta_n)} g_{n_1-\frac{1}{2}}(\nu_1-\ii\eta_m)g_{n_2-\frac{1}{2}}(\nu_2-\ii\eta_n)\frac{s(\frac{5}{2}-\nu_1-\ii\eta_m,\lambda)s(\frac{5}{2}-\nu_2-\ii\eta_n,\lambda)}{[s(2,\lambda)]^2}\right],
\label{eq:cov_binave_2d}
\end{align}
\end{widetext}
\noindent where $B$ is the denominator of Eq.~(\ref{eq:cov_binave_2d_def}),
\begin{equation}
    B =\frac{1}{4}\left(\bar{\theta}_{i,\rm max}^2-\bar{\theta}_{i,\rm min}^2\right)\left(\bar{\theta}_{j,\rm max}^2-\bar{\theta}_{j,\rm min}^2\right)~,
\end{equation}
$\lambda$ is again the constant logarithmic bin width defined by $\bar{\theta}_{i,{\rm max}}/\bar{\theta}_{i,{\rm min}}=e^\lambda$. $\ell_0$ is the smallest sampled angular wave number and $\theta_0$ is the smallest sampled angular separation. $n_1,n_2$ are the orders of (cylindrical) Bessel functions in Eq.~(\ref{eq:3by2cov}).

In Eq.~(\ref{eq:cov_binave_2d}), we have assumed a logarithmically uniform angular binning. For a general choice of binning, we must directly compute Eq.~(\ref{eq:cov_binave_2d_general}). The complexities are similar to the $D=3$ case discussed in \S\ref{subsec:binave}.

\section{Validation for Future Survey Analyses}\label{sec:validation}
We implement the 2D-FFTLog algorithm into the covariance module of the \textsc{CosmoLike} analysis framework \citep{2014MNRAS.440.1379E,2017MNRAS.470.2100K}. To validate the angular bin-averaged flat sky 3$\times$2pt covariance computed with the 2D-FFTLog algorithm, we also implement an angular bin-averaged flat sky covariance using a direct quadrature (Quad) integration, and an angular bin-averaged curved sky covariance (detailed in \S\ref{subsubsec:covs}). The curved sky covariance is the most accurate among the three by definition.

We compute the three 3$\times$2pt covariances for a DES Y3-like and an LSST Y1-like survey. Negligible differences are found between the flat sky covariances and the curved sky covariances. To confirm that the differences do not impact on the cosmological constraints, we perform simulated 3$\times$2pt likelihood analyses using Monte Carlo Markov Chains (MCMC) with these covariances.

\subsection{Analysis Ingredients}\label{subsec:ingredients}
\subsubsection{Lens and Source Galaxy Sample Distributions}\label{subsubsec:samples}
We assume a DES Y3-like survey covering 5000 deg$^2$. All other analysis settings are adopted from the DES Y1 analysis \citep{2018PhRvD..98d2006E,2018MNRAS.481.2427C}. Specifically, we assume a \textsc{RedMaGiC} \citep{2016MNRAS.461.1431R} lens sample, split into 5 tomographic bins. The galaxy number density in each bin is [0.0134, 0.0343, 0.0505, 0.0301, 0.0089] per arcmin$^{2}$, and the fiducial linear galaxy bias parameters are set as [1.44, 1.70, 1.70, 2.00, 2.06]. For the source sample, we assume a \textsc{Metacalibration} selected sample \citep{2017arXiv170202600H,2017ApJ...841...24S,2018MNRAS.481.1149Z}, split into 4 bins again similar to the DES Y1 analysis \citep{2018MNRAS.478..592H,2018PhRvD..98d3528T}. The source galaxy number density in each bin is [1.496, 1.5189, 1.5949, 0.7949] per arcmin$^{2}$. We also assume shape noise consistent with DES Y1, \ie, $\sigma_\epsilon=0.279$ per ellipticity component.

For LSST Y1, we generate the redshift distributions of the lens and source galaxies following the DESC Science Requirements Document \citep[DESC SRD,][]{2018arXiv180901669T}. Our LSST Y1-like survey has a survey area of 12300 deg$^2$, and is expected to measure galaxies with an i-band limit of 24.1 mag for the large-scale structure (lens sample) and an i-band depth 25.1 mag for the weak lensing (source sample). For the lens sample, we use a parametric redshift distribution consistent with the DESC SRD, \ie, $dN/dz\propto z^2\exp[-(z/z_0)^\alpha]$, with $(z_0,\alpha)=(0.26,0.94)$, normalized by the effective number density $n_{\rm eff}=18\,$arcmin$^{-2}$. We then split the sample into 5 equally populated tomographic bins (different from the DESC SRD) based on the estimated redshifts (including photo-$z$ errors), and convolve each bin with a Gaussian photo-$z$ scatter with $\sigma_z = 0.03(1+z)$ as an estimate for the bin’s true redshift distribution. We set the fiducial linear galaxy bias parameter for each bin as $b_i = 1.05/G(\bar{z}^i)$, where $\bar{z}^i$ is the mean redshift of the $i$-th bin, and $G(z)$ is the linear growth factor. For the source sample, we use the same parametric form but with $(z_0,\alpha)=(0.191,0.870)$, normalized to $n_{\rm eff}=11.2\,$arcmin$^{-2}$.\footnote{These values for the source sample are the updated version from private communication with Rachel Mandelbaum.} We also split the source sample into 5 equally populated tomographic bins and convolve each bin with a Gaussian photo-$z$ uncertainty with $\sigma_z=0.05(1+z)$. The distributions of the LSST Y1 lens and source tomographic bins are the same as in \cite{2019arXiv191111947F}, and are shown on the left panel of Fig.~4 in \cite{2019arXiv191111947F}. We assume the galaxy shape noise to be $\sigma_\epsilon= 0.26$ per component.

\subsubsection{Angular Two-Point Functions}\label{subsubsec:2ptfuncs}
We model the galaxy-galaxy lensing (GGL) and the cosmic shear power spectra using the Limber approximation. Including redshift space distortion (RSD), lensing magnification (Mag), galaxy intrinsic alignment (IA), and following the notation in \cite{2019arXiv191111947F}, the GGL power spectrum between lens bin $i$ and source bin $j$ can be written as
\begin{equation}
    C_\ell^{{\rm g}^ie^j} = \frac{2}{2\ell+1}\int_0^\infty dk\,\tilde{\Delta}^{{\rm g}^i}(\chi_\ell)\tilde{\Delta}^{e^j}(\chi_\ell)P_\delta(k,z(\chi_\ell))~,
\end{equation}
and the tomographic cosmic shear power spectrum between the source bin $i$ and $j$ can be written as
\begin{equation}
    C_\ell^{e^ie^j} = \frac{2}{2\ell+1}\int_0^\infty dk\,\tilde{\Delta}^{e^i}(\chi_\ell)\tilde{\Delta}^{e^j}(\chi_\ell)P_\delta(k,z(\chi_\ell))~,
\end{equation}
where $\ell$ is the angular wavenumber, $k$ is the wavenumber of the Fourier mode. $P_\delta(k,z)$ is the nonlinear matter power spectrum at redshift $z$. The argument $\chi_\ell = \ell/k$ arises from the Limber approximation. $\tilde{\Delta}^{\rm g},\tilde{\Delta}^{\rm e}$ are the modified transfer functions of the galaxy number overdensity $\delta_{\rm g}$ and the observed galaxy shape $e$, respectively, following the definitions in Eqs.~(4.3-4.5) and (4.12-4.13) in \cite{2019arXiv191111947F} \citep[also see][]{2018arXiv181205995C}, \ie,
\begin{align}
    \tilde{\Delta}^{\rm g}=& \tilde{\Delta}^{\rm D}(\chi_\ell)+\frac{3\ell(\ell+1)\Omega_{\rm m} H_0^2(1+z(\chi_\ell))}{c^2k^2}W^{\rm M}(z(\chi_\ell))\nonumber\\
    &+\frac{1+8\ell}{(2\ell+1)^2}\frac{f(z(\chi_\ell))}{b(z(\chi_\ell))}\tilde{\Delta}^{\rm D}(\chi_\ell)\nonumber\\
    &-\frac{4}{2\ell+3}\sqrt{\frac{2\ell+1}{2\ell+3}}\frac{f(z(\chi_{\ell+1}))}{b(z(\chi_{\ell+1}))}\tilde{\Delta}^{\rm D}(\chi_{\ell+1})~,
\end{align}
\begin{align}
    \tilde{\Delta}^{e} = \sqrt{\frac{(\ell+2)!}{(\ell-2)!}}&\left[\frac{3\Omega_{\rm m} H_0^2(1+z(\chi_\ell))}{2c^2k^2}W^{\rm L}(z(\chi_\ell))\right.\nonumber\\
    &\ \left.+\frac{n_{\rm src}(z(\chi_\ell))A_{\rm IA}(z(\chi_\ell))H(z(\chi_\ell))}{c(\ell+1/2)^2}\right]~,
\end{align}
where $\tilde{\Delta}^{\rm D}(\chi_\ell)=\frac{1}{c}n_{\rm lens}(z(\chi_\ell))b(z(\chi_\ell))H(z(\chi_\ell))$, $n_{\rm lens}(z)$ is the redshift distribution of the lens galaxies, $b(z)$ is the linear galaxy bias parameter, $f(z)$ is the logarithmic growth factor, $H(z)$ is the Hubble parameter, and $c$ is the speed of light. $W^{\rm M}(z)$ and $W^{\rm L}(z)$ are the lensing magnification kernel function and the lensing kernel function, defined as
\begin{align}
    W^{\rm M}(z) &=\int_z^{\infty} dz' n_{\rm lens}(z')\frac{b_{\rm mag}(z')}{2}\frac{\chi(z')-\chi(z)}{\chi(z)\chi(z')}~,\\
    W^{\rm L}(z)&=\int_z^\infty dz'\,n_{\rm src}(z')\frac{\chi(z')-\chi(z)}{\chi(z')\chi(z)},
\end{align}
where $\chi(z)$ is the comoving distance at redshift $z$. $b_{\rm mag}(z)$ is the magnification bias parameter encapsulating the linear dependence of the galaxy number density on the convergence $\kappa$ at a given point on the sky, defined such that the galaxy overdensity $\delta_{\rm g}$ is changed by $\Delta\delta_{\rm g}=b_{\rm mag}\kappa\delta_{\rm g}$. $n_{\rm src}(z)$ is the redshfit distribution of the source galaxies. We adopt the ``nonlinear linear alignment model'' of IA \cite[\eg,][but see \citet{2017arXiv170809247B} for limitation]{2001MNRAS.320L...7C,2004PhRvD..70f3526H,2007MNRAS.381.1197H,2011A&A...527A..26J,2015PhR...558....1T,2015JCAP...08..015B,2016MNRAS.456..207K}. $A_{\rm IA}(z)$ is the (dimensionless) alignment amplitude, defined by
\begin{equation}
    A_{\rm IA} = -\frac{C_1\rho_{\rm cr}\Omega_{\rm m}}{G(z)}a_{\rm IA}\left(\frac{1+z}{1+z_0}\right)^{\eta}~,
\end{equation}
where we use $C_1\rho_{\rm cr}\simeq 0.0134$, a normalization derived from SuperCOSMOS observations \citep{2004PhRvD..70f3526H,2007NJPh....9..444B}, $z_0$ is an arbitrary pivot value for the power-law scalings of the redshift (with index parameter $\eta$), which we take $z_0=0.62$ in our analysis, following the DES Year 1 choice in \cite{2018PhRvD..98d3528T}. We have reduced the number of free parameters by absorbing the luminosity dependence into the free parameter $a_{\rm IA}$.

The Limber approximation may induce significant parameter shifts when applied to the galaxy clustering auto power spectra $C_\ell^{{\rm g}^i{\rm g}^i}$ for future DES and LSST analyses as shown in \cite{2019arXiv191111947F}. Therefore, we model these power spectra without the Limber approximation, and adopt the method in \S4.1 of \cite{2019arXiv191111947F}.

We compute the linear matter power spectrum using the transfer function from \cite{1998ApJ...496..605E}, and the nonlinear matter power spectrum with \textsc{HaloFit} \citep{2003MNRAS.341.1311S,2012ApJ...761..152T}. We calculate the angular two-point correlation functions for galaxy clustering $w^i(\theta)$, GGL $\gamma_t^{ij}(\theta)$, and cosmic shear $\xi_{+/-}^{ij}(\theta)$, using their relation to angular power spectra on the curved sky \citep[\eg][]{1996astro.ph..9149S,2010PhRvD..82j3522D}:
\begin{align}
    w^i(\theta) &= \sum_\ell \frac{2\ell+1}{4\pi}P_\ell(\cos\theta) C_\ell^{{\rm g}^i{\rm g}^i}~,\\
    \gamma_t^{ij}(\theta) &= \sum_\ell \frac{2\ell+1}{4\pi\ell(\ell+1)}P^2_\ell(\cos\theta) C_\ell^{{\rm g}^i e^j}~,\\
    \xi_{\pm}^{ij}(\theta) &= \sum_\ell\frac{2\ell+1}{2\pi\ell^2(\ell+1)^2}[G_{\ell,2}^+(\cos\theta)\pm G_{\ell,2}^-(\cos\theta)] C_\ell^{e^i e^j}~,
\end{align}
where $\theta$ is the angular separation, $P_\ell$ and $P_\ell^2$ are the Legendre polynomial and the associated Legendre polynomial, $G_{\ell,m}^{+/-}$ are given by Eq.~(4.19) of \cite{1996astro.ph..9149S}.

For DES Y3, we compute all correlation functions in 20 logarithmically spaced angular bins over the range $2.5'<\theta<250'$, while for LSST Y1, we compute all correlation functions in 26 logarithmically spaced angular bins over $2.5'<\theta<900'$. For each angular bin $[\theta_{\rm min},\theta_{\rm max}]$, the correlation functions are bin-averaged, \ie, replacing $P_\ell(\cos\theta)$, $P_\ell^2(\cos\theta)$ and $[G_{\ell,2}^+(\cos\theta)\pm G_{\ell,2}^-(\cos\theta)]$ with their bin-averaged functions $\overline{P_\ell}$, $\overline{P^2_\ell}$ and $\overline{G_{\ell,2}^+\pm G_{\ell,2}^-}$ (Friedrich et al., in preparation), defined by Eqs.~(5.6-5.8) in \cite{2019arXiv191111947F}.
Note that this curved sky bin-averaging reduces to the flat sky bin-averaging when $\theta$ is small. Our analysis includes all auto-correlations of the lens bins for the galaxy clustering, all combinations of lens and source bins for the GGL, and all auto- and cross-correlations of the source bins for the cosmic shear. Thus, for DES Y3, the data vector contains 5 sets of $w(\theta)$, 20 sets of $\gamma_t(\theta)$, 10 sets of $\xi_+(\theta)$ and 10 sets of $\xi_-(\theta)$, each of which has 20 angular bins. For LSST Y1, the data vector contains 5 sets of $w(\theta)$, 25 sets of $\gamma_t(\theta)$, 15 sets of $\xi_+(\theta)$ and 15 sets of $\xi_-(\theta)$, each of which has 26 angular bins.

\subsubsection{Systematics}\label{subsubsec:systematics}
We parametrize systematic uncertainties through a set of nuisance parameters closely following the DES Y1 analysis \citep{2017arXiv170609359K}. We also add the lensing magnification effect to the modeling, and marginalize over its amplitude. The fiducial values and the prior distributions of the parameters are summarized in Table \ref{tab:params}.
\paragraph*{Photometric redshift uncertainties} The uncertainty in the redshift distribution of the $i$-th tomographic bin $n^i(z)$ is modeled by one shift parameter $\Delta_z$ for each bin of the lens and the source samples, \ie, $n^i(z) = \hat{n}^i(z-\Delta_z^i)$, where the index $i$ traverses over all the lens and source bins, and $\hat{n}$ is the estimated redshift distribution as described in \S\ref{subsubsec:samples}. There are 9 shift parameters in total for DES Y3 and 10 shift parameters for LSST Y1. We take 0 as their fiducial values to generate the simulated data vector, and marginalize over them in the likelihood analyses. For the DES Y3 lens sample, we choose a Gaussian prior with $\mu=0,\sigma=[4, 3, 3, 5, 11]\times 10^{-3}$ for each $\Delta_{z,\rm lens}^i$; for the source sample, we choose a Gaussian prior with $\mu=0,\sigma=0.005$ for each $\Delta_{z,\rm source}^i$,
For LSST Y1, we choose a Gaussian prior with $\mu=0,\sigma=0.005(1+\bar{z}^i)$ for each $\Delta_{z,\rm lens}^i$, and a Gaussian prior with $\mu=0,\sigma=0.002(1+\bar{z}^i)$ for each $\Delta_{z,\rm source}^i$, consistent with the requirements given in \S5.1 and 5.2 of the DESC SRD.

\paragraph*{Galaxy bias} We assume a linear bias model and use one parameter for each lens bin. There are 4 parameters in total for DES Y3, and 5 for LSST Y1, whose fiducial values are described in \S\ref{subsubsec:samples} for generating the simulated data vector. In the likelihood analysis, they will be marginalized over with conservative flat priors [0.8, 3].

\paragraph*{Lensing magnification bias} We parameterize the lensing magnification effect through one parameter for each lens bin $b_{\rm mag}^i$ (see \S\ref{subsubsec:2ptfuncs} for modeling details). For magnitude limited samples, magnification due to lensing by line-of-sight structure can affect the number density of galaxies with observed magnitudes exceeding the magnitude cut \citep[\eg,][]{1995astro.ph.12001V,1998MNRAS.294L..18M,2008PhRvD..77b3512L}. For LSST Y1, we assume the lens samples are magnitude limited, and estimate $b_{\rm mag}^i =$ ($-0.898$, $-0.659$, $-0.403$, $-0.0704$, $0.416$) for the 5 lens bins as in \cite{2019arXiv191111947F}, which is based on the fitting formula in \cite{2010A&A...523A...1J} and $r$-band limit given by the DESC SRD. For DES Y3 lens sample, we adopt the DES Year 6 $b_{\rm mag}$ values, $b_{\rm mag}^i =$ ($-0.102$, $-0.102$, $-0.102$, $1.06$, $1.06$), that we estimate in \cite{2019arXiv191111947F}, which is based on the Schechter luminosity function \citep{1976ApJ...203..297S} and the luminosity cuts $L/L_*>(0.5,0.5,0.5,1.0,1.0)$ for bins from low to high redshifts, respectively. $L_*$ is the characteristic galaxy luminosity where the power-law form in the Schechter luminosity function cuts off.
The 5 nuisance parameters $b_{\rm mag}^i$ are marginalized over conservative flat priors [$-3$,3] in the likelihood analysis.

\paragraph*{Multiplicative shear calibration} We use one shear calibration uncertainty parameter $m^i$ per source bin (4 in total for DES Y3, and 5 for LSST Y1), acting on the cosmic shear and GGL correlation functions such that
\begin{equation}
    \xi_{+/-}^{ij}(\theta) \rightarrow (1+m^i)(1+m^j)\xi_{+/-}^{ij}(\theta)~,~~
    \gamma_t^{ij}(\theta) \rightarrow (1+m^j)\gamma_t^{ij}(\theta)~.
\end{equation}
The $m^i$ are marginalized over independently with Gaussian priors ($\mu=0,\sigma=0.005$).

\paragraph*{IA} We use the nonlinear linear alignment (NLA) model and parameterize it with two parameters $a_{\rm IA}$ and $\eta$ (see \S\ref{subsubsec:2ptfuncs} for modeling details). Their fiducial values are $a_{\rm IA}=0.5$ and $\eta=0$, and they are both marginalized over independently with conservative flat priors [$-5$, 5].

\subsubsection{Covariances}\label{subsubsec:covs}
The implementation of the Fourier space covariance matrices is detailed in the Appendix of \cite{2017MNRAS.470.2100K}. The computation of the bin-averaged flat sky real space covariances is introduced in \S\ref{sec:3by2pt}, where the ``flat sky'' approximation is applied to both the bin-averaging step and the transform from Fourier to real space covariance. For comparison, we also compute the covariances of bin-averaged correlation functions on the curved sky (without the flat sky approximation in either step), \ie, for two angular two-point functions, $\Xi,\Theta\in\lbrace w,\gamma_t,\xi_+,\xi_-\rbrace$
\begin{equation}
    \cov(\Xi^{ij}(\bar{\theta}),\Theta^{km}(\bar{\theta}')) = \sum_\ell\overline{P^\Xi_\ell}(\bar{\theta})\sum_{\ell'}\overline{P^\Theta_{\ell'}}(\bar{\theta}')\,\cov(C_\Xi^{ij}(\ell),C_\Theta^{km}(\ell'))~,
\label{eq:cov_fullsky}
\end{equation}
where $C_{\xi_+}= C_{\xi_-}= C^{ee}$, $C_{\gamma_t}= C^{{\rm g}e}$, and $C_w= C^{\rm gg}$ in our previous notation, and $i,j,k,m$ are the tomographic bin indices. The bin-averaged weight functions are defined as (Friedrich et al., in preparation)
\begin{align}
    &\overline{P^w_\ell}= \frac{2\ell+1}{4\pi}\overline{P_\ell}~,~~\overline{P^{\gamma_t}_\ell}= \frac{2\ell+1}{4\pi\ell(\ell+1)}\overline{P^2_\ell}~,\nonumber\\
    &\overline{P^{\xi_{\pm}}_\ell}= \frac{2\ell+1}{2\pi\ell^2(\ell+1)^2}\overline{G_{\ell,2}^+\pm G_{\ell,2}^-}~.
\end{align}
In our implementation, we evaluate Eq.~(\ref{eq:cov_fullsky}) up to $\ell_{\rm max}=50000$. Thus, the transform has complexity of order $\mathcal{O}(\ell_{\rm max}^2 N_\theta^2)$, where $N_\theta=20$ is the number of angular bins.
\begin{figure*}
    \centering
    \includegraphics[width=0.7\textwidth]{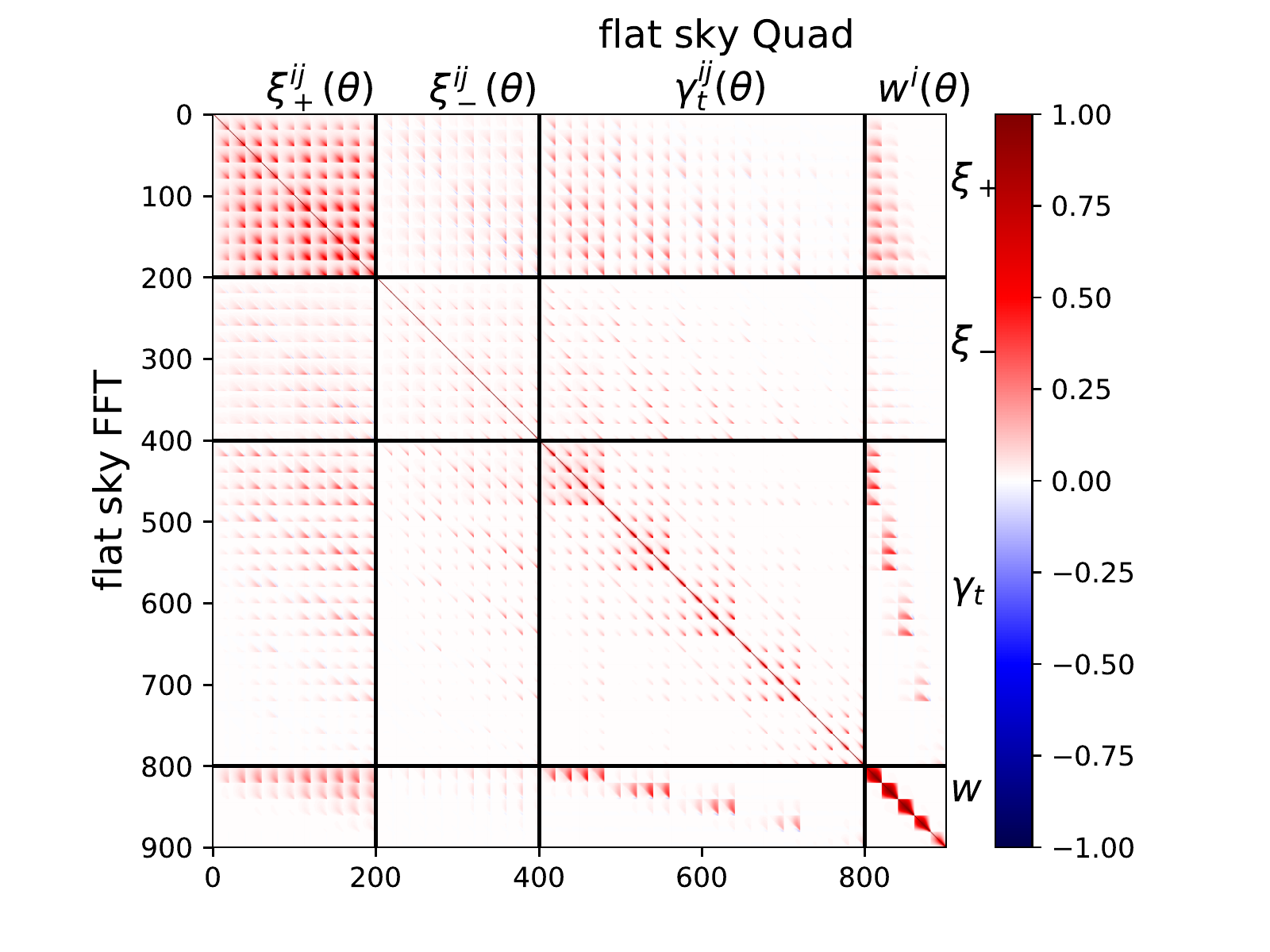}
    \caption{DES Y3 3$\times$2pt real space flat sky correlation matrices from our FFT method (lower triangle) and the direct quadrature (Quad) integration (upper triangle). The matrices are 900$\times$ 900, and the blocks of covariances between different probes are annotated.}
    \label{fig:prec_mat}
\end{figure*}

\subsubsection{Angular Scale Cuts}\label{subsubsec:scalecuts}
Survey analyses define a set of angular scale cuts to prevent nonlinear modeling limitations and baryonic feedback processes from biasing the cosmology results. For LSST Y1, we choose a scale cut $k_{\rm max}=0.3h/$Mpc as in the DESC SRD, which roughly corresponds to a comoving scale $R_{\rm min}=2\pi/k_{\rm max}=21\,$Mpc$/h$. For the galaxy clustering, we define the angular scale cut 
$\theta_{\rm min}^{w^i}$ for lens tomographic bin $i$ as $\theta_{\rm min}^{w^i} = R_{\rm min}/\chi(\bar{z}^i)~$, where $\bar{z}^i$ is the mean redshift of galaxies in tomographic bin $i$. For the lens sample, $\theta_{\rm min}^{w^i}$ are [$80.88'$, $54.19'$, $42.85'$, $35.43'$, $29.73'$]. The scale cuts for $\gamma_t^{ij}$ are the same as $\theta_{\rm min}^{w^i}$. For the cosmic shear, we use cuts $\ell<\ell_{\rm max}=3000$ as defined in the DESC SRD, and translate it into the angular cuts for $\xi_{+/-}$ with the first zeros of their corresponding Bessel functions $J_{0/4}$ (in the flat sky-limit transform), \ie, $\theta_{\rm min}^{\xi_+}=2.4048/\ell_{\rm max}=2.756'$, and $\theta_{\rm min}^{\xi_-}=7.5883/\ell_{\rm max}=8.696'$.

For DES Y3, we adopt $R_{\rm min}=8\,$Mpc$/h$ for the galaxy clustering, and $R_{\rm min}=12\,$Mpc$/h$ for the GGL, same as DES Y1 analysis choice \citep{2017arXiv170609359K}. For cosmic shear, we adopt scale cuts same as those used in DES Y1 cosmic shear analysis \citep{2018PhRvD..98d3528T}.

\subsection{Covariance Comparison}\label{subsec:cov_compare}
We implement three versions of the 3$\times$2pt angular bin-averaged covariances: (1) the flat sky covariance using the 2D-FFTLog algorithm, (2) the flat sky covariance using a direct quadrature integration, and (3) the curved sky covariance introduced in \S\ref{subsubsec:covs}.

We first compare these covariances at high level (\eg, eigenvalues and singular values, as well as their signal-to-noise ratios in Fig.~\ref{fig:snr}) in \S\ref{subsubsec:eigens}. Then, we estimate the impact of the numerical artifacts in the covariances on the inferred goodness of fit ($\chi^2$) in \S\ref{subsubsec:chi2}. Finally in \S\ref{subsubsec:like}, we test the impact on the inferred means and uncertainties of the cosmological parameters using simulated likelihood analyses. The first two tests are very easy to perform and useful to check before running the likelihood analysis.

\subsubsection{Elements, Eigenvalues and Singular Values}\label{subsubsec:eigens}
\begin{figure*}
    \centering
    \includegraphics[width=0.49\textwidth]{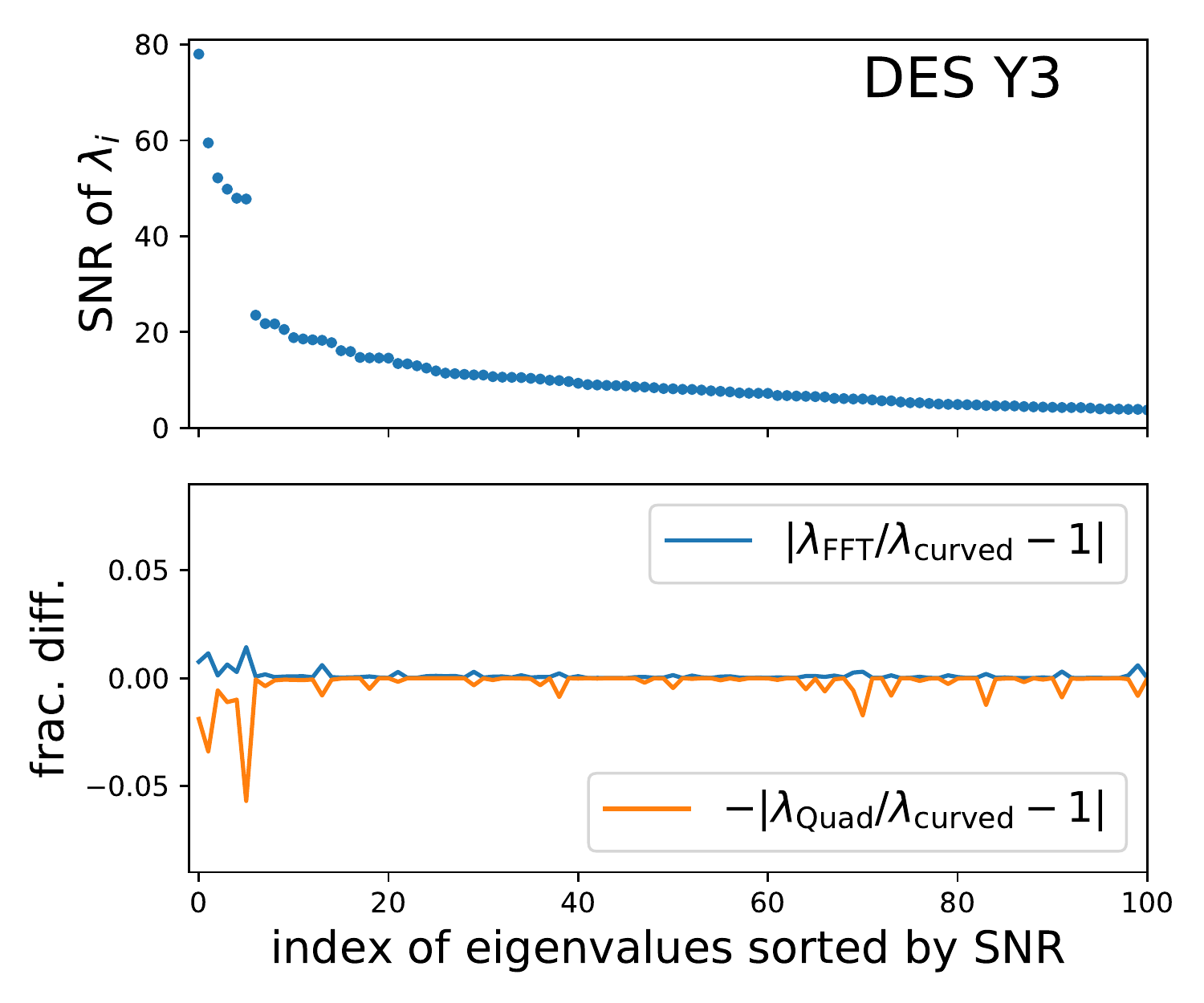}\includegraphics[width=0.49\textwidth]{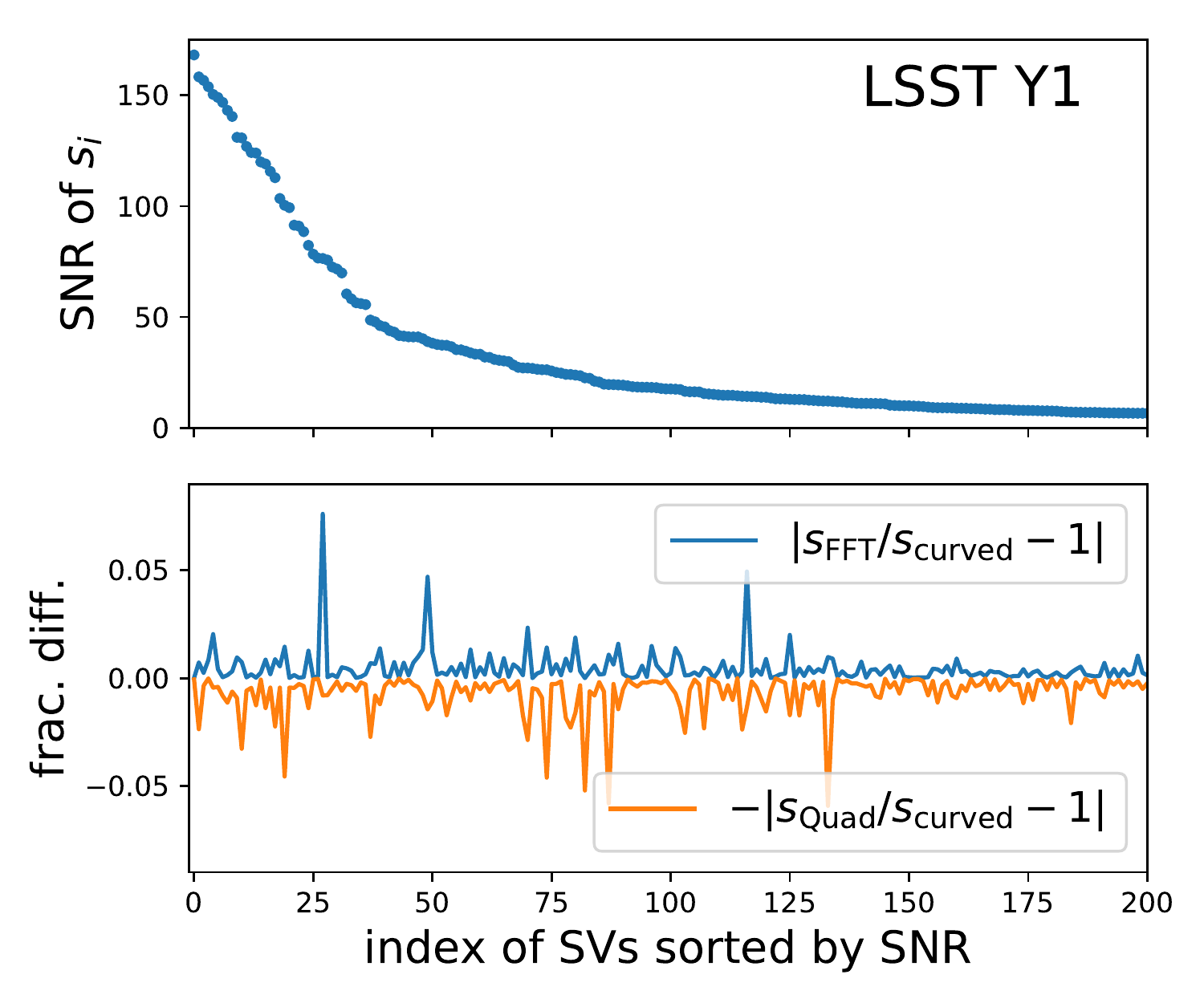}
    \caption{The SNRs of eigenvalues of the DES Y3 curved sky covariance (upper left) and the SNRs of SVs of the LSST Y1 curved sky covariance (upper right), sorted by magnitude. The eigenvalues and SVs are also compared to the flat sky FFT and Quad covariances, respectively (lower left and right). For each survey, the flat sky eigenvalues and SVs match the curved sky eigenvalues and SVs within a few percent. We only show the 100 (out of 900) eigenvalues of the highest SNRs for DES Y3, and 200 (out of 1560) SVs for LSST Y1.}
    \label{fig:snr}
\end{figure*}
We provide several high level comparisons of the three versions of 3$\times$2pt angular bin-averaged covariances in terms of their elements, eigenvalues and singular values.

For a visual element-wise comparison, Fig.~\ref{fig:prec_mat} shows the 900$\times$900 DES Y3 3$\times$2pt real space flat sky correlation matrices $C_{ij}/\sqrt{C_{ii}C_{jj}}$ from FFT and Quad methods, where no obvious difference is detectable by eye.

We then compute the eigenvalues of the three DES Y3 covariances and the three LSST Y1 covariances, and further compute the signal-to-noise ratio (SNR) of each eigenvalue. The total SNR is defined as ${\rm SNR}^2 = \mathbfit{D}^{\rm T}\mathbfss{C}^{-1}\mathbfit{D}$, where $\mathbfit{D}$ is the fiducial 3$\times$2pt data vector consisting the 3 types of angular two-point correlation functions (see \S\ref{subsubsec:2ptfuncs}). Since $\mathbfss{C}$ is a real symmetric matrix, the eigendecomposition of $\mathbfss{C}$ leads to
\begin{equation}
    \mathbfss{C} = \mathbfss{V}\mathbf{\Lambda}\mathbfss{V}^{\rm T}~,
\end{equation}
where $\mathbf{\Lambda}$ is the diagonal matrix of all the eigenvalues $\lambda_i$ of $\mathbfss{C}$ sorted in descending order, and $\mathbfss{V}$ is an orthogonal matrix ($\mathbfss{V}^{-1}=\mathbfss{V}^{\rm T}$) whose columns are the eigenvectors of $\mathbfss{C}$. Thus, we have
\begin{equation}
    {\rm SNR}^2 = \sum_{i}\frac{(\mathbfss{V}^{\rm T}\mathbfit{D})_i^2}{\lambda_i} = \sum_i {\rm SNR}_i^2~,
\end{equation}
where we have defined the SNR of the $i$-th eigenvalue $\lambda_i$ in the last equality.

As the flat sky Quad LSST Y1 covariance suffers from numerical instabilities with 1 negative eigenvalue, we do not include it in further analyses. Instead, we perform the singular value (SV) decomposition of LSST Y1 covariances, and compute the SNR of each SV. We decompose $\mathbfss{C}$ as $\mathbfss{C}=\mathbfss{U}\bm{\Sigma}\mathbfss{V}^{\rm T}$, where $\mathbfss{U},\mathbfss{V}$ are both orthogonal, and $\bm{\Sigma}$ is a diagonal matrix of all the SVs $s_i$ sorted in descending order. Similar to the eigendecomposition, we have
\begin{equation}
    {\rm SNR}^2 = \sum_{i}\frac{(\mathbfss{U}^{\rm T}\mathbfit{D})_i(\mathbfss{V}^{\rm T}\mathbfit{D})_i}{s_i} = \sum_i {\rm SNR}_i^2~,
\end{equation}
where we define the SNR for the $i$-th SV $s_i$ in the last equality.

Fig.~\ref{fig:snr} shows a comparison of the eigenvalues and SVs of the flat and curved sky covariances. We sort the SNRs of the 900 eigenvalues of the DES Y3 covariances and the 1560 SVs of the LSST Y1 covariances, and then compare the eigenvalues/SVs of flat sky covariances to those of the curved sky covariances. For each survey, the eigenvalues/SVs show few-percent level agreement, and we only show the 100 eigenvalues/200 SVs of the highest SNRs. Overall, the flat sky FFT covariances show better agreement with the curved sky covariance, especially for those eigenvalues/SVs with high SNRs. This may attribute to lower numerical accuracy of the Quad calculations, and may result in the negative eigenvalue of the LSST Y1 flat sky Quad covariance.

\subsubsection{Shifts in $\chi^2$}\label{subsubsec:chi2}
An inaccurate covariance matrix affects the inferred goodness of fit. The goodness of fit may be quantified by $\chi^2$ in the data vector space, defined as $\chi^2=(\mathbfit{D}-\mathbfit{M}(\mathbfit{p}))^{\rm T}\mathbfss{C}^{-1}(\mathbfit{D}-\mathbfit{M}(\mathbfit{p}))$, where $\mathbfit{D}$ is the 3$\times$2pt data vector, $\mathbfit{M}(\mathbfit{p})$ is the model vector evaluated with cosmological and nuisance parameters $\mathbfit{p}$, and $\mathbfss{C}$ is the covariance matrix.

We assume that the curved sky covariance is the true covariance $\mathbfss{C}_0$, and that $\mathbfit{x}=\mathbfit{D}-\mathbfit{M}(\mathbfit{p})$ follows a Gaussian distribution, $\mathbfit{x}\sim N(0,\mathbfss{C}_0)$. Using a ``false'' covariance $\mathbfss{C}_1$ will result in a shift in $\chi^2$, \ie, $\Delta\chi^2=\mathbfit{x}^{\rm T}(\mathbfss{C}_1^{-1}-\mathbfss{C}_0^{-1})\mathbfit{x}$. Both $\mathbfit{x}^{\rm T}\mathbfss{C}_1^{-1}\mathbfit{x}$ and $\mathbfit{x}^{\rm T}\mathbfss{C}_0^{-1}\mathbfit{x}$ are quadratic forms. We compute the expectation value and the variance of $\Delta\chi^2$ per degree of freedom (d.o.f.),
\begin{align}
    E[\Delta\chi^2/N_D] &= {\rm tr}(\mathbfss{C}_1^{-1}\mathbfss{C}_0)/N_D - 1~,\\
    {\rm Var}[\Delta\chi^2/N_D] &= [2N_D + 2{\rm tr}(\mathbfss{C}_1^{-1}\mathbfss{C}_0\mathbfss{C}_1^{-1}\mathbfss{C}_0)-4{\rm tr}(\mathbfss{C}_1^{-1}\mathbfss{C}_0)]/N_D^2~,
\end{align}
where $N_D$ is the dimension of the data vector (900 for DES Y3, 1560 for LSST Y1, and significantly less when scale cuts are applied).

Table \ref{tab:chisq_shift} shows the means and standard deviations of $\Delta\chi^2$ per d.o.f. if the flat sky FFT or Quad covariances are used for DES Y3 and LSST Y1, respectively. The $\chi^2$ shift per d.o.f. with scale cuts tends to be larger due to the larger weights of the large angular scales. This $\chi^2$ shift is the $\chi^2$ difference due to the change of covariance matrix only, and does not account for the possible shift in the best-fit values of parameters, which can happen when data vectors are fitted with inaccurate covariances. Given the insignificance of these $\chi^2$ shifts, there is reason to believe that the impact of using the flat sky covariances for DES Y3 and LSST Y1 on the goodness of fit is negligible. However, we note that the corresponding $\chi^2$ shifts for fitting noisy data vectors can be larger and needs to be checked on a case by case basis.

\begin{table*}
\footnotesize
    \centering
    \begin{tabular}{|c|c|c|c|}
    \hline
    \multirow{2}{*}{Surveys} &\multirow{2}{*}{d.o.f.} & \multicolumn{2}{ c| }{$\Delta\chi^2$ per d.o.f.} \\ \cline{3-4}
     & & FFT vs curved & Quad vs curved \\
    \hline
    DES Y3 no scale cuts & 900& $0.00074\pm 0.00041$ & $0.00174\pm 0.00058$\\
    \ \ \ \ \ \ \ \ \,with scale cuts & 447& $0.00160\pm 0.00075$ & $0.00313\pm 0.00101$\\
    \hline
    LSST Y1 no scale cuts & 1560& $0.00159\pm 0.00154$ & -- \\
    \ \ \ \ \ \ \ \ \ \,with scale cuts & 1059& $0.00359\pm 0.00208$ & -- \\
    \hline
    \end{tabular}
    \caption{The $\chi^2$ shifts per d.o.f. of using the flat sky FFT/Quad covariances for DES Y3-like and LSST Y1-like surveys, assuming that the curved sky covariances are true. Both surveys with and without scale cuts are considered. The flat sky Quad covariance of LSST Y1 is not included as it is not invertible in our implementation. The insignificance of these $\chi^2$ shifts indicates negligible impact of using the flat sky covariances for DES Y3 and LSST Y1 on the goodness of fit.}
    \label{tab:chisq_shift}
\end{table*}

\subsubsection{Simulated Likelihood Analysis}\label{subsubsec:like}

An inaccurate covariance ultimately affects the inferred cosmological parameter constraints. We investigate this impact by performing simulated likelihood analyses.

We generate the simulated 3$\times$2pt data vector $\mathbfit{D}$ by computing the model vector at the fiducial parameter values and in our fiducial cosmology, \ie, the standard $\Lambda$CDM with massless neutrinos, with non-Limber modeling of $w$ and $\gamma_t$ \citep[see \S4.2 of][for modeling details of $\gamma_t$]{2019arXiv191111947F}, and Limber modeling of $\xi_{+/-}$ \citep[see \eg,][for discussion on the impact of Limber approximation on weak lensing]{2017JCAP...05..014L}. Throughout our analyses, we use the \textsc{emcee} sampler \citep{2013PASP..125..306F}. The fiducial values and priors of the parameters (30 parameters are sampled for DES Y3, and 32 for LSST Y1) are summarized in Table \ref{tab:params}. We focus on the cosmological parameter subspace ($\Omega_m,\sigma_8,n_s$), since galaxy clustering and weak lensing are most powerful in constraining $\Omega_m$ and $\sigma_8$, while $n_s$ is a good indicator of whether the model has sufficiently corrected the parameter biases due to the Limber approximation \citep{2019arXiv191111947F}.

\begin{figure*}
    \centering
    \includegraphics[width=0.49\textwidth]{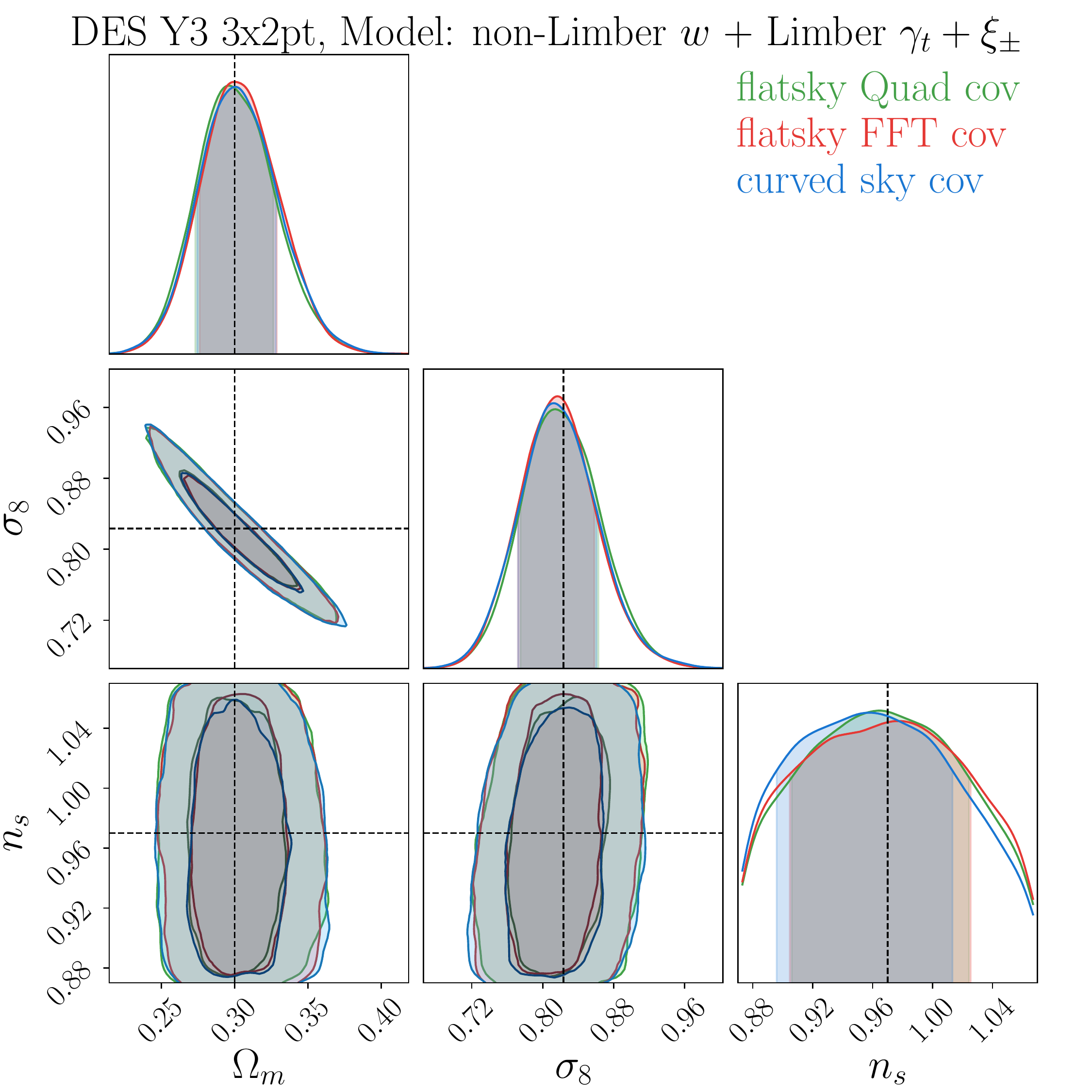}
    \includegraphics[width=0.49\textwidth]{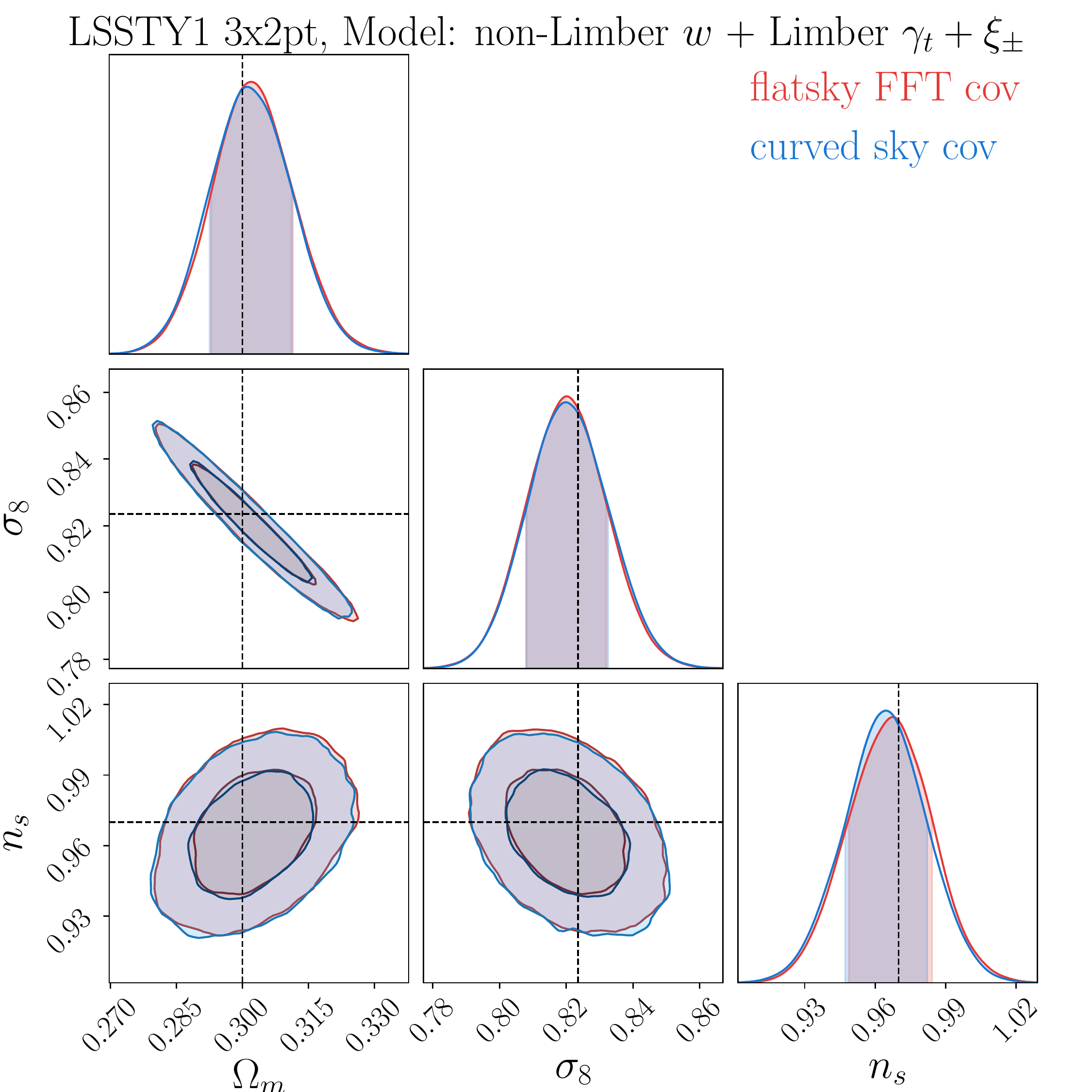}
    \caption{The 1$\sigma$ and 2$\sigma$ contours of fitting the simulated 2x2pt data vector with the model, using bin-averaged flat sky covariances and the curved sky covariances for DES Y3 (\textit{left}) and LSST Y1 (\textit{right}). For DES Y3, we have two versions of flat sky covariances from the direct quadrature (Quad) integration and our 2D-FFTLog algorithm. For LSST Y1, we only compare the FFT result to the curved sky result.}
    \label{fig:contours}
\end{figure*}

The left panel of Fig.~\ref{fig:contours} shows the constraints on the 3 cosmological parameters from simulated DES Y3 analyses with the three different covariances. They all recover the fiducial values of the cosmological parameters with nearly identical constraints. The right panel shows a comparison of the cosmological constraints from simulated LSST Y1 analyses with the two different covariances, and we again find good agreement of the inferred best fit parameters and their uncertainties. We conclude that the 2D-FFTLog algorithm is sufficiently accurate, and that using the flat sky limit of the 3$\times$2pt covariance up to 900 arcmins angular separation will not bias the cosmological parameters ($\Omega_m,\sigma_8,n_s$) for an LSST Y1-like 3$\times$2pt analysis.

\begin{table*}
\footnotesize
    \centering
    \begin{tabular}{ |l|c|c| }
    \hline
    Parameters & Fiducial & Prior \\
    \hline
    \textbf{Survey} & & \\
    $\Omega_{\rm survey}$ & DES 5000 deg$^2$; LSST 12300 deg$^2$ & fixed \\
    $\sigma_e$ per component & DES 0.279; LSST 0.26 & fixed \\
    \hline
    \textbf{Cosmology} & & \\
    $\Omega_{\rm m}$ & 0.3 & flat [0.1, 0.9]\\
    $\sigma_8$ & 0.82355 & flat [0.4, 1.2] \\
    $n_{\rm s}$ & 0.97 & flat [0.87, 1.07]\\
    $\Omega_b$ & 0.048 & flat [0.03, 0.07]\\
    $h_0$ & 0.69 & flat [0.55, 0.91] \\
    $w_0$ & -1 & fixed \\
    $w_a$ & 0 & fixed \\
    $\sum m_\nu$
    & 0 & fixed \\
    \hline
    \textbf{Galaxy Bias} & & \\
    $b^i$ & DES [1.44, 1.70, 1.70, 2.00, 2.06]; & flat [0.8, 3]\\
     &  LSST [1.24, 1.36, 1.47, 1.60, 1.76] & \\
    \hline
    \textbf{Magnification Bias} & & \\
    $b_{\rm mag}^i$ & DES [-0.102, -0.102, -0.102, 1.06, 1.06]; & flat [-3, 3]\\
     & LSST [-0.898, -0.659, -0.403, -0.0704, 0.416] & \\
    \hline
    \textbf{Lens/Source Photo-$z$} & & \\
    $\Delta_{z,\rm lens}^i$ & 0 & LSST Gauss $(0, 0.005(1+\bar{z}^i_{\rm lens}))$;\\
    &  & DES Gauss $(0,[4,3,3,5,11]\times 10^{-3})$\\
    $\Delta_{z,\rm source}^i$ & 0 & LSST Gauss $(0, 0.002(1+\bar{z}^i_{\rm src}))$;\\
    &  & DES Gauss $(0, 0.005(1+\bar{z}^i_{\rm src}))$\\
    \hline
    \textbf{Shear Calibration} & & \\
    $m^i$ & 0 & Gauss (0, 0.005)\\
    \hline
    \textbf{IA} & & \\
    $a_{\rm IA}$ & 0.5 & flat [-5, 5]\\
    $\eta$ & 0 & flat [-5, 5]\\
    \hline
    \end{tabular}
    \caption{A list of the parameters characterizing the surveys, cosmology and systematics. The entries are separated by a semi-colon if they are different for DES Y3 and LSST Y1; otherwise, we only write out the shared entry. The fiducial values are used for generating the simulated data vector, and the priors are used in the sampling. Flat priors are described by [minimum, maximum], and Gaussian priors are described by Gauss ($\mu, \sigma$).}
    \label{tab:params}
\end{table*}

\section{Discussion and Summary}
\label{sec:discussion}
Statistical inference of cosmological parameters for future surveys require precise covariance matrices. The covariance matrices can be estimated from data or an ensemble of simulations, or calculated analytically. Computing covariances analytically is much faster and less noisy. However, the transform of covariance matrices from Fourier space to real space involves integrals with two Bessel integrals (for 3D statistics or projected statistics at flat sky limit), which are numerically unstable due to the oscillatory nature of the integrand. These numerical issues are more severe for large angular separation of the correlation functions and for lower noise (\ie, for higher number density of galaxies or lower shape noise of weak lensing shear measurements), leading to longer computation time. The issue is exacerbated for higher dimensional covariance matrices, as the computation time grows quadratically with the dimension.

We present a 2D-FFTLog algorithm to compute the real space bin-averaged Gaussian and non-Gaussian covariances (\S\ref{sec:algorithm}). The algorithm is as accurate as traditional methods, but much faster, with complexity of order $\mathcal{O}(N^2\log N)$, where $N$ is the size of the sampled Fourier space covariance. In contrast, traditional methods are of order $\mathcal{O}(N_k^2 N_r^2)$, where $N_k\gg N$ is the required number of sampling points in Fourier space, and $N_r$ is the size of the desired real space covariance. We apply the algorithm to the covariances of angular correlation functions of galaxy clustering, galaxy-galaxy lensing, and cosmic shear (\S\ref{sec:3by2pt}), and validate our method for DES Y3-like and LSST Y1-like surveys (\S\ref{sec:validation}). For both surveys, we compare the flat sky FFT covariance to the exact, but computationally slow, curved sky transformation, the ``curved sky covariance'', and find that the flat sky FFT covariance is sufficiently accurate and does not bias cosmological parameters, even at LSST Y1 precision.

In terms of the speed, for a 20$\times$20 bin-averaged covariance block, 2D-FFTLog takes $\sim 1\,$second on one CPU to perform the transform with a sampling size $N=10^3$ (and additional zero-padding), while the ``curved sky covariance'' takes $0.5-1$ hour. The computation time of the quadrature integration varies significantly from seconds to a few hours depending on the orders of the Bessel functions, with longer time for higher orders and cross-covariances with beating between different Bessel function orders.

The 2D-FFTLog algorithm can also be applied to other real space covariances of projected statistics, such as 5x2pt, 6x2pt \citep[joint analyses of 3$\times$2pt and CMB lensing,][]{2017PhRvD..95l3512S,2019PhRvD..99b3508B,2019PhRvD.100b3541A}, and cluster clustering and lensing, where the method in \S\ref{sec:3by2pt} is directly applicable, as well as covariances of 3D statistics, such as multipoles of galaxy 3D correlation functions \citep[\eg,][]{2016MNRAS.457.1577G}, where the method in \S\ref{sec:algorithm} is directly applicable.

The 2D-FFTLog code, written in both python and C, is publicly available at \url{https://github.com/xfangcosmo/2DFFTLog}. The C code is incorporated into \textsc{CosmoLike} \citep{2017MNRAS.470.2100K}. We release an improved version of the real-space 3$\times$2pt covariance code \textsc{CosmoCov} that we have built for this paper at \url{https://github.com/CosmoLike/CosmoCov}.

\section*{Acknowledgments}
We thank an anonymous referee for detailed comments and suggestions. We thank Oliver Friedrich, Niall MacCrann, Jonathan Blazek, Marko Simonovi\'c, Vivian Miranda for helpful discussions and comments. We thank Oliver Friedrich and Stella Seitz for their distributed notes on bin-averaging and curved sky covariances, and Rachel Mandelbaum for the updated LSST Y1 source sample distribution. We also thank Shivam Pandey and Anna Porredon for testing \textsc{CosmoCov}, and thank Felipe Andrade-Oliveira for pointing out a typo in an equation.

XF, TE are supported by NASA ROSES ATP 16-ATP16-0084 grant. TE is supported by Department of Energy grant DE-SC0020215. EK is supported by Department of Energy grant DE-SC0020247. Calculations in this paper use High Performance Computing (HPC) resources supported by the University of Arizona TRIF, UITS, and RDI and maintained by the UA Research Technologies department.

\bibliographystyle{mnras}
\bibliography{references.bib}


\bsp	
\label{lastpage}
\end{document}